\documentclass[journal]{IEEEtran}
\IEEEoverridecommandlockouts

\usepackage{comment}
\usepackage{caption}
\usepackage{stfloats} 
\usepackage{booktabs}
\usepackage{diagbox}
\usepackage{tabularx}
\usepackage{multirow} 
\usepackage{amsmath}
\usepackage[boxed,commentsnumbered,ruled,linesnumbered]{algorithm2e}
\usepackage[justification=centering]{caption}
\usepackage[ruled,linesnumbered]{algorithm2e}

\usepackage{varwidth}
\usepackage{amsmath,amssymb,amsfonts,graphicx}

\usepackage{hyperref}
\hypersetup{colorlinks=true,citecolor = blue}
\usepackage{graphicx,epstopdf,caption,url} 
\usepackage{amssymb} 
\usepackage{array} 
\usepackage{arydshln} 
\usepackage{graphics}
\usepackage{color}  
\usepackage{graphicx}  

\newtheorem{definition}{Definition} 
\newtheorem{theorem}{\textbf{Theorem}}
\newtheorem{strategy}{\textbf{Strategy}}

\begin{document}

\title{UPER: Efficient Utility-driven Partially-ordered \\Episode Rule Mining}

\author{Hong Lin, Wensheng Gan*, Junyu Ren, Philip S. Yu,~\IEEEmembership{Life Fellow,~IEEE} 

\thanks{This research was supported in part by National Natural Science Foundation of China (No. 62272196), the Guangzhou Basic and Applied Basic Research Foundation (No. 2024A04J9971). Hong Lin and Wensheng Gan contributed equally to this work.}  

\thanks{Hong Lin, Wensheng Gan, and Junyu Ren are with the College of Cyber Security, Jinan University, Guangzhou 510632, China.  (E-mail:  lhed9eh0g@gmail.com, wsgan001@gmail.com, renjunyu193@gmail.com)} 

\thanks{Philip S. Yu is with the Department of Computer Science, University of Illinois Chicago, Chicago, USA. (E-mail: psyu@uic.edu)} 
    
\thanks{Corresponding author: Wensheng Gan}
}

\maketitle

\begin{abstract}
    Episode mining is a fundamental problem in analyzing a sequence of numerous events. For discovering strong relationships between events in a complex event sequence, episode rule mining is proposed. However, both the episode and episode rules have strict requirements for the order of events. Hence, partially-ordered episode rule mining (POERM) is designed to loosen the constraints on the ordering, i.e., events in the antecedents and consequents of the rule can be unordered, and POERM has been applied to real-life event prediction. In this paper, we consider the utility of POERM, intending to discover more valuable rules. We define the utility of POERs and propose an algorithm called UPER to discover high-utility partially-ordered episode rules. In addition, we adopt a data structure named NoList to store the necessary information, analyze the expansion of POERs in detail, and propose several pruning strategies (namely WEUP, REUCSP, and REEUP) to reduce the number of candidate rules. Finally, we conduct experiments on several datasets to demonstrate the effectiveness of different pruning strategies in terms of time and memory optimization.
\end{abstract}


\IEEEpeerreviewmaketitle

\section{Introduction}

Pattern mining (aka pattern discovery) \cite{fournier2022pattern,gan2019survey} is useful for discovering meaningful information and insights from big data. Over the past few decades, many algorithms have been proposed to analyze sequential data, where a sequence refers to a list consisting of events or symbols with an order \cite{fournier2022maxfem}. For example, in consumer purchase behavior analysis, the sequence $<$apple, milk, bread$>$ represents that apples are purchased first, followed by milk, and finally bread. Sequential pattern mining (SPM) aims at finding patterns in multiple discrete sequences. It has been extensively studied, including frequency-driven sequential pattern mining \cite{agrawal1995mining, fournier2014fast}, utility-driven sequential pattern mining \cite{yin2012uspan, lan2014applying}, frequency-driven sequential rule mining \cite{fournier2011rulegrowth, fournier2014erminer}, utility-driven sequential rule mining \cite{zida2015efficient, huang2023us}, etc. Besides, there is a type of task that also deals with sequential data called episode pattern mining. Different from SPM, the data processed by episode pattern mining is a single sequence with time stamps and consisting of multiple discrete events \cite{fournier2022maxfem}. One of the most common tasks is frequent episode pattern mining (FEM), which aims to discover all frequent episodes. Mannila \textit{et al.} \cite{mannila1997discovery} first proposed two FEM algorithms named MINEPI and WINEPI. To improve efficiency, Huang \textit{et al.} \cite{huang2008efficient} proposed two algorithms named MINEPI+ and EMMA. Since too many episodes may be mined when the support threshold is set too low, TKE \cite{fournier2020tke} was proposed to discover the top-$k$ most frequent episodes. In summary, the sequence processed by FEM can be divided into two types: simple event sequence and complex event sequence, and the difference is whether events are allowed to occur simultaneously at the same time point in the event sequence. Although frequency-driven pattern mining can efficiently discover those frequent patterns, it may miss information that is valuable but infrequent. For example, consider applying FEM to analyze user behavior risk over a period on a website and classifying user behavior into low-risk behaviors (e.g., page visit, login) and risky behaviors (e.g., SQL injection, uploading Trojan files). In this scenario, FEM discovers mostly frequent but low-risk behaviors and ignores high-risk behaviors of concern. ``Utility" is used to describe the importance of an item or an event. For instance, in consumer purchase behavior analysis, the profit of an item is its utility, and in the above analysis of user website behavior, the risk of the behavior can be considered its utility. Wu \textit{et al.} first \cite{wu2013mining} defined high-utility episode pattern mining (HUEM) and proposed a novel algorithm to discover HUEs. HUEM is more challenging than FEM due to two reasons. On the one hand, the utility of a pattern does not exhibit anti-monotonicity, unlike frequency. Therefore, upper bounds with anti-monotonicity have been proposed and improved to enhance mining efficiency, such as \textit{Episode Weighted Utilization} (\textit{EWU}) \cite{wu2013mining}, \textit{Optimized EWU} (\textit{EWU}$_{\textit{opt}'}$) \cite{gan2023discovering}, and \textit{Episode-Remaining Utilization} (\textit{ERU}) \cite{fournier2019hue}. On the other hand, when there are multiple occurrences of an event, traditional HUEM tends to miss some high-utility episodes because of the constraints imposed by following a specific processing order \cite{fournier2019hue}. Thus, it remains challenging to comprehensively consider the utility of all occurrences of the episode. HUEM has a wide range of applications, including stock prediction \cite{lin2017novel}, cross-marketing \cite{rathore2016top}, and workload prediction in the cloud \cite{amiri2018online}.

Since episode patterns that occur frequently in databases with low confidence are not valuable for prediction or decision-making, another type of pattern called episode rules has been proposed. Episode rule mining is an extension of episode pattern mining, and it can reveal strong temporal relationships between events due to the consideration of confidence. Several algorithms with different frequency measures have been proposed, including PPER \cite{ao2017mining} using minimal occurrences, NONEPI \cite{ouarem2021mining} using non-overlapping occurrences, etc. To extract more informative episode rules, Lin \textit{et al.} \cite{lin2015discovering} first considered the discovery of high-utility episode rules in a complex event sequence. However, for a rule $r$: $X$ $\rightarrow$ $Y$, they only treat the utility of $Y$ as the utility of $r$, while ignoring the utility of $X$. It may lead to ignoring the weights of $X$ in some event prediction scenarios. Besides, the standard episode rule is not general enough since it is totally-ordered \cite{chen2021sequence}. For example, a standard episode rule $r_1$: \{login\} $\rightarrow$ \{page visit, SQL injection, uploading Trojan files\} indicates that behaviors page visit and SQL injection occur after login, and SQL injection occurs after page visit, and uploading Trojan files occurs after SQL injection. This means that other similar rules, such as $r_2$: \{login\} $\rightarrow$ \{page visit, uploading Trojan files, SQL injection\}, $r_3$: \{login\} $\rightarrow$ \{page visit, SQL injection, uploading Trojan files\} and $r_4$: \{login\} $\rightarrow$ \{uploading Trojan files, page visit, SQL injection\} will be treated differently. However, when only the order of the antecedents $X$ and consequents $Y$ of the rule is concerned by the user, these rules may represent the same situation as someone who logs in first and then performs the rest of the behavior. At this point, analyzing a multitude of rules that represent the same situation but with slightly different ordering is inconvenient for the user. To loosen the constraints on the ordering of events in the antecedents and consequents of the episode rule, Fournier-Viger \textit{et al.} \cite{fournier2021mining} first defined the partially-ordered episode rule (\textit{POER}) and proposed a formal framework called POERM. The events in the antecedents and consequents of POERs are unordered; thus, a partially-ordered episode rule can represent multiple traditional episode rules. For instance, the rule $r_1$ can replace other episode rules $r_2$, $r_3$, and $r_4$. POERM measures support for episode rules using non-overlapping occurrences. Subsequently, Chen \textit{et al.} \cite{chen2021mining} adopted the header support frequency to measure the support. They also conducted experiments on different datasets and concluded that partially-ordered episode rules usually provide more accurate predictions than standard episode rules \cite{chen2021sequence}. In addition, several studies have applied POERM to prediction tasks in different areas, such as regulatory activities \cite{jarvela2023predicting} and smartphone context events \cite{goyal2023smartphone}. 

To the best of our knowledge, no research on mining high-utility episode rules is based on partially-ordered sets and considers the utility of both $X$ and $Y$. It is crucial to fill this gap. For example, the rule $r_1$: \{login\} $\rightarrow$ \{page visit, SQL injection, uploading Trojan files\} contains high-risk (utility) behaviors that occur more rarely than normal visitor behaviors. If the utility value is not considered, such important partially ordered episode rules will be ignored. In this paper, we propose a novel mining approach named UPER (\underline{U}tility-driven \underline{P}artially-ordered \underline{E}pisode \underline{R}ule mining) to consider utility in POERM. To summarize, there are four key contributions in this paper, as follows:

\begin{itemize}	
    \item  We define a novel type of the high-utility partially-ordered episode rule. This type of pattern loosens the ordering constraints on the antecedents and consequents of an episode rule and fully accounts for their utility.
    
    \item  We develop an effective algorithm named UPER with a novel data structure named NoList for storing rich information about the occurrence of event sets and rules, including position and utility value.

    \item We propose several upper bounds and pruning strategies, aimed at reducing the search space and mining partially-ordered high-utility episode rules more efficiently. 

    \item We perform several experiments to analyze the impact of different thresholds on the efficiency of the algorithm, and the results show the advantages of the proposed pruning strategies in runtime and memory consumption.
\end{itemize}

The remaining parts of this paper are given as follows. Related work is stated and summarized in Section \ref{sec: relatedwork}. A series of preliminaries and basic knowledge are described in Section \ref{sec: preliminaries}. The UPER algorithm with its pruning strategies is detailed in Section \ref{sec: algorithm}. Furthermore, the experimental results are shown in Section \ref{sec: experiments}. Finally, we present the conclusion and future work in Section \ref{sec: conclusion}.

\section{Related Work} \label{sec: relatedwork}

\subsection{Frequent Episode Mining}

Pattern mining has been a vital research area in data knowledge discovery in recent decades. Sequential pattern mining (SPM) \cite{fournier2017survey, gan2019survey} is one of the most extensive research areas, aiming at discovering interesting knowledge in a sequential database, where the sequence of the database is composed of items in order. For example, frequency-based SPM, utility-driven SPM \cite{gan2019survey}, frequent sequential rule mining \cite{fournier2014erminer}, non-overlapping SPM \cite{wu2017nosep, wu2020netncsp}, etc. Different from sequence mining, episode mining focuses on a sequence of discrete events that occur at different points in time (time points) in the sequence. An episode consists of multiple events that exist in a time order, which can be applied to represent events that occur during a certain period of time in real life. In the past, many studies about frequency-based episode mining were proposed, such as MINEPI and WINEPI \cite{mannila1997discovery}, EMMA and MINEPI+ \cite{huang2008efficient}. It is worth noting that, because an episode may have multiple occurrences in a sequence, there are different methods to measure frequency. The most commonly used measure is the minimal occurrence \cite{mannila1997discovery, meger2004constraint}, in addition to non-overlapping occurrences \cite{laxman2005discovering}, head frequency \cite{huang2008efficient}, sliding windows \cite{mannila1997discovery}, etc. Among them, Huang \textit{et al.} \cite{huang2008efficient} indicated that window-based support may provide more interesting patterns than non-overlapping occurrences. Ouarem \textit{et al.} \cite{ouarem2023discovering} defined a new concept called distinct occurrences to compute the frequency of episodes and proposed the EMDO algorithm for mining parallel episodes and episode rules. In general, the way the episode is extended in FEM can be categorized into two types: parallel extension and serial extension. Parallel extension refers to the extension of those sets of events that occur simultaneously with the episode, while serial extension refers to the extension of those sets of events that occur after the episode. Moreover, a common challenge with FEM algorithms is poor scalability when dealing with large-scale data. Therefore, Ao \textit{et al.} \cite{ao2019large} proposed a scalable distributed framework to address this challenge.

\subsection{Utility-driven Episode Mining}

In general, the discovered results are mainly based on the frequency measure. Utility-driven pattern mining (UPM) \cite{gan2021huopm,gan2021survey,gan2021fast} is a specialized area within the discipline of pattern mining that focuses on the discovery of high-utility patterns within datasets. In the field of UPM, there are different types of discovered results based on the types of dataset, including high-utility itemsets in transaction data \cite{nguyen2023efficient, kim2022ehmin}, high-utility sequential patterns in sequence data \cite{zhang2023mining,gan2021fast}, high-utility sequential rules in sequence data \cite{huang2023us}, and high-utility subgraphs in a graph \cite{alam2023ugmine}. In addition, UPM in dynamic databases has been widely studied. For example, Baek \textit{et al.} \cite{baek2021rhups} recently proposed a novel algorithm to discover recent high-utility patterns from stream databases. However, all these approaches do not address the complete event sequences, which means that they do not consider both the counts of the event appearances (i.e., internal utility) and the weights of the events (i.e., external utility) at any point in time. Wu \textit{et al.} \cite{wu2013mining} first considered the utility factor in episode mining, defined the high-utility episode (\textit{HUE}), and proposed the UP-Span method with an upper bound called Episode Weighted Utilization (\textit{EWU}). To discover \textit{HUEs} faster, Guo \textit{et al.} \cite{guo2014high} proposed the algorithm TSpan, which improves efficiency by using the lexicographic prefix tree. They also use several strategies to reduce the search space of the tree. Gan \textit{et al.} \cite{gan2023discovering} proposed an efficient mining algorithm named UMEpi (Utility Mining of high-utility Episodes from complex event sequences) by utilizing the improved \textit{EWU} calculation (called \textit{EWU}$_{opt'}$), which is tighter than the original \textit{EWU}. Note that UMEpi is a depth-first way to directly mine HUEs without scanning the sequence multiple times. The HUE-Span algorithm \cite{fournier2019hue} proposed a more compact upper bound, namely Episode-Remaining Utilization (\textit{ERU}). It pointed out that previous studies were not accurate in calculating episode utility and focused on improving the calculation by considering the maximal utility for minimal occurrences. In short, episode pattern mining has been extensively studied, but it fails to reveal the relationship between two episodes, so a new task named episode rule mining has been proposed.

\subsection{Episode Rule Mining}

Episode rule mining is an extension of episode pattern mining that aims at discovering strong relationships between events in a complex event sequence \cite{fournier2019efficient}. Similar to FEM, there are different measures of the frequency of episode rules. PPER \cite{ao2017mining} uses a minimal occurrence and sliding-window-based approach to find episode rules. Ouarem \textit{et al.} \cite{ouarem2021mining} proposed the NONEPI algorithm by considering non-overlapping occurrences to measure frequency for the first time. To discover more meaningful episode rules, Lin \textit{et al.} \cite{lin2015discovering} combined utility into episode rules and proposed a direct way to generate high-utility episode rules. The algorithm, called UBER-Mine, improves performance with the help of a compact tree structure, the UR-Tree. However, they only consider the utility of the $Y$ as the utility of the rule, which may result in the weight of $X$ being ignored in some event prediction scenarios. Besides, the mining task has strict requirements on the order of occurrence of events in $X$ and $Y$, whose results are not partially-ordered. The standard episode rule is totally-ordered, and it constrains the order so strictly that it is not general enough \cite{chen2021sequence}. To loosen the constraints on the ordering of events in the antecedents and consequents of the episode rules, Fournier-Viger \textit{et al.} \cite{fournier2021mining} firstly defined the partially-ordered episode rule (\textit{POER}) and proposed a formal framework called POERM, which first discovers all frequent single events and then expands them to discover all antecedents of the rule (denoted as $X$). Subsequently, POERM composes $X$ with all single events to form rule $X$ $\rightarrow$ \{$e$\}, and then expands the consequents of rule (denoted as $Y$) to get all valid \textit{POERs} in the form of $X$ $\rightarrow$ $Y$. Chen \textit{et al.} \cite{chen2021mining} presents an algorithm called POERMH, which uses the head support frequency measure for more accurate sequence prediction. The above motivational reasons inspired us to develop new algorithms. In this paper, we consider the utility of partially-ordered episode rule mining to loosen the constraints of total ordering and discover more valuable episode rules.

\section{Preliminary and Problem Statement}  
\label{sec: preliminaries}

\begin{definition}[Complex event sequence]
   \rm Let $E$ = \{$e_1$, $e_2$, $\dots$, $e_l$\} be a set of $l$ distinct events. A pair ($e$, $T_i$) means that the event $e$ occurs in a time stamp $T_i$. For each $e$ that occurs at each $T_i$ there is a corresponding value called \textit{utility}, which indicates the importance of $e$. A complex event sequence $S$ consists of all event sets, i.e., $S$ = $<$(\textit{SE}$_{1}$, $T_1$), (\textit{SE}$_{2}$, $T_2$), \dots, (\textit{SE}$_{n}$, $T_n$)$>$, where \textit{SE}$_i$ represents the set of all events occurring in $T_i$ simultaneously. In particular, when only one event occurs at each time point of a sequence, it is called a simple event sequence.
\end{definition}

\begin{figure}[h]
    \centering
    \includegraphics[trim=0 0 0 0,clip,scale=0.4]{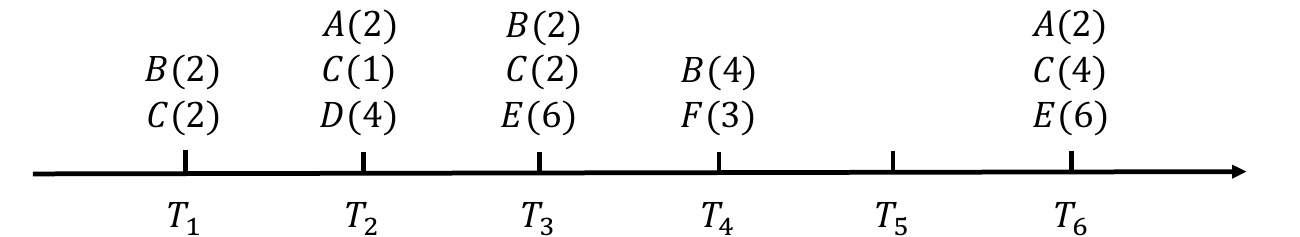}
    \caption{A complex event sequence.}
    \label{fig:sequence}
\end{figure}

As shown in Fig. \ref{fig:sequence}, there is a complex event sequence $S$ = $<$((\textit{BC}), $T_1$), ((\textit{ACD}), $T_2$), ((\textit{BCE}), $T_3$), ((\textit{BF}), $T_4$), ((\textit{ACE}), $T_6$)$>$ with utility since there exists no less than one event occurring at each time point. Besides, note that there is no event at time point $T_5$, thus $T_5$ is not represented in $S$.

\begin{definition}[Event set and partially-ordered episode rule \cite{fournier2021mining}]
    \rm  An event set $F$ is a subset of $E$, and we call it a \textit{k}-event set if it contains \textit{k} events. In this paper, events in an event set can be either serial or simultaneous. An episode rule $r$: $X$ $\rightarrow$ $Y$ is a relationship between two event sets $X$ and $Y$, which indicates that events in $Y$ may occur after all the events in $X$ have occurred. A partially-ordered episode rule means that there is no order between events in $X$ and $Y$. Besides, we refer to the rule whose $X$ and $Y$ both have only one event, as the \textit{1*1 rule}. For convenience, the rule mentioned in this paper refers to the partially-ordered episode rule.
\end{definition}

\begin{definition}[Occurrence and interval]
   \rm If an event $e$ in $S$ occurs at more than one time point, it is regarded as having multiple occurrences. Therefore, an event set $F$ or an episode rule $r$ may also have multiple occurrences. More precisely, we describe different occurrences of $F$ or $r$ by the interval. The interval [$T_i$, $T_j$] of $F$ occurrence means that events in $F$ occur earliest in $T_i$ and latest in $T_j$, while [$T_i$, $T_j$] of $r$ occurrence means that events in $X$ occur earliest in $T_i$ and the events in $Y$ occur latest in $T_j$. The set of all occurrences of $e$, $F$ and $r$ is denoted as \textit{occ}($e$), \textit{occ}($F$) and \textit{occ}($r$), respectively.
\end{definition}

\begin{definition}[Time duration constraint \cite{fournier2021mining}]
\label{de:time constraint}
   \rm  For an episode rule $r$: $X$ $\rightarrow$ $Y$, three user-specified positive integer constants, \textit{XSpan}, \textit{YSpan}, and \textit{XYSpan}, are used to avoid the events it contains being too far apart in time. Assuming that an occurrence of $r$ and its intervals of $X$ and $Y$ are [\textit{X.start}, \textit{X.end}] and [\textit{Y.start}, \textit{Y.end}], it is valid when it satisfies: 1) \textit{X.end} $-$ \textit{X.start} $<$ \textit{XSpan}, 2) \textit{Y.end} $-$ \textit{Y.start} $<$ \textit{YSpan}, 3) \textit{Y.start} $-$ \textit{X.end} $<$ \textit{XYSpan}. Hence, we can get \textit{Y.end} $-$ \textit{X.start} $\leq$ \textit{WinSpan} and \textit{WinSpan} = \textit{XSpan} + \textit{YSpan} + \textit{XYSpan} $-$ 3, and the proof is as follows:
\end{definition}

Since \textit{XSpan}, \textit{YSpan}, and \textit{XYSpan} are all positive integers, we can get \textit{X.end} $-$ \textit{X.start} $\leq$ \textit{XSpan} $-$ 1, \textit{Y.end} $-$ \textit{Y.start} $\leq$ \textit{YSpan} $-$ 1 and \textit{Y.start} $-$ \textit{X.end} $\leq$ \textit{XYSpan} $-$ 1. In summary, \textit{Y.end} $-$ \textit{X.start} $\leq$ \textit{XSpan} + \textit{YSpan} + \textit{XYSpan} $-$ 3.

\begin{definition}[Overlapping and non-overlapping occurrences]
   \rm For any two occurrences [$T_{i1}$, $T_{j1}$] and [$T_{i2}$, $T_{j2}$] of an event set $F$ or a rule $r$, if $T_{i2}$ $\leq T_{i1}$ $\leq T_{j2}$ or $T_{i1} \leq T_{i2} \leq T_{j1}$, then these two occurrences are overlapping with each other. Conversely, if $T_{j1}$ $<$ $T_{i2}$ or $T_{j2}$ $<$ $T_{i1}$ holds, then the two occurrences are non-overlapping. Besides, we denote the set of all non-overlapping occurrences of $F$ and $r$ as \textit{nocc}($F$) and \textit{nocc}($r$), respectively.
\end{definition}

For example, if \textit{XSpan} = \textit{YSpan} = \textit{XYSpan} = 3, for the event set $F$ = \{$B$, $F$\}, its occurrences are \{[\textit{T$_3$}, $T_4$], [\textit{T$_4$}, $T_4$]\} but the size of \textit{nooc}($F$) is 1. The rule $r$ = \{$B$\} $\rightarrow$ \{$C$\} has three occurrences, i.e., \textit{occ}($r$) = \{[$T_1$, $T_2$], [$T_1$, $T_3$], [$T_4$, $T_6$]\}, but the size of \textit{nooc}($r$) is 2.

\begin{definition}[Support and confidence of an episode rule \cite{fournier2021mining}]
   \rm For an episode rule $r$ = $X \rightarrow Y$, its support is the number of all non-overlapping occurrences of $r$. We denote it as \textit{sup}($r$) = $|$\textit{nocc}($X \rightarrow Y$)$|$. Besides, the confidence of $r$ is defined as \textit{conf}($r$) = $|$\textit{nocc}($X \rightarrow Y$)$|$$/$$|$\textit{nocc}($X$)$|$, and it means the proportion of $Y$ occurs after $X$.
\end{definition}

For example, \textit{sup}(\{$B$\} $\rightarrow$ \{$C$\}) = 2 and \textit{conf}(\{$B$\} $\rightarrow$ \{$C$\}) = \textit{sup}(\{$B$\} $\rightarrow$ \{$C$\})$/$\textit{sup}(\{$B$\}) = 2$/$3 = 0.67.

\begin{definition}[Utility of an event set]
    \rm For an occurrence [$T_i$, $T_j$] of an event set $F$, its utility value is equal to the sum of the utilities of occurrences of all events in $F$, i.e., $u$($F$, [$T_i$, $T_j$]) = $\sum_{e \in F}$${u(e, [T_i, T_j])}$. The utility of $F$ is the sum of the utilities of all non-overlapping occurrences, i.e., $u$($F$) = $\sum_{F \in nocc(F)}{u(F, [T_i, T_j])}$. For those overlapping occurrences, the maximum utility among them is defined as their utility.
\end{definition}

\begin{definition}[Utility of an episode rule]
   \rm We denote the utility of an occurrence [$T_i$, $T_j$] of a rule $r$: $X$ $\rightarrow$ $Y$ as $u$($r$, [$T_i$, $T_j$]), and its value is equal to the sum of the utilities of the antecedent event set and the consequent event set, i.e., $u$($r$, [$T_i$, $T_j$]) = $u$($X$, [$T_i$, $T_m$]) + $u$($Y$, [$T_n$, $T_j$]) where the occurrence of $X$ is [$T_i$, $T_m$], the occurrence of $Y$ is [$T_n$, $T_j$] and $i$ $\leq$ $m$ $<$ \textit{n} $\leq$ $j$. Then the utility of $r$ is the sum of the utilities of all non-overlapping occurrences, i.e., $u$($r$) = $\sum_{r \in nocc(r)}{u(r, [T_i, T_j])}$. For those overlapping occurrences, the maximum utility among them is defined as their utility.
\end{definition}

For instance, in Fig. \ref{fig:sequence}, $u$(\{$B$, $C$\}, [$T_1$, $T_1$]) = 4 and $u$(\{$B$, $C$\}) = $u$(\{$B$, $C$\}, [$T_1$, $T_1$]) + $u$(\{$B$, $C$\}, [$T_3$, $T_3$]) = 8 if \textit{XSpan} = 1. For the rule $r$ = \{$B$\} $\rightarrow$ \{$C$\}, $u$($r$) = \textit{max}\{$u$($r$, [$T_1$, $T_2$]), $u$($r$, [$T_1$, $T_3$])\} + $u$($r$, [$T_4$, $T_6$]) = 4 + 8 = 12 if \textit{XSpan} = \textit{YSpan} = 1 and \textit{XYSpan} = 3.

\textbf{Problem definition.} Given a utility-based event sequence $S$, three time duration constraints (\textit{XSpan}, \textit{YSpan} and \textit{XYSpan}), minimum support threshold (\textit{minsup}), minimum confidence threshold (\textit{minconf}), and a minimum utility threshold (\textit{minutil} = $\delta$ $\times$ $u$($S$)). The task aims at mining all the high-utility episode rules in a complex event sequence $S$. A rule $r$ is defined as a high utility episode rule (denoted as \textit{HUER}) if it satisfies: \textit{HUER} $\leftarrow$ \{$r$ $|$ $u$($r$) $\geq$ \textit{minutil} $\wedge$ \textit{sup}($r$) $\geq$ \textit{minsup} $\wedge$ \textit{conf}($r$) $\geq$ \textit{minconf}\}.

\section{The UPER Algorithm}  \label{sec: algorithm}

We introduce the details of the UPER algorithm in this section. We first analyzed the change in position before and after the expansion of the episode rules. Subsequently, we propose several upper bounds and pruning strategies and give proofs. We also introduce the data structures used by the algorithm and describe the procedure in detail.

\subsection{Definition and Pruning Strategies}

\begin{definition}[Rule expansion]
\label{de:expansion}
    \rm In this paper, we use the same rule-growth method as the \textit{POERM} to expand a rule $r$ = $X$ $\rightarrow$ $Y$, referred to as \textit{Y expansion}. For an expanded event $e$, the \textit{Y expansion} of $r$ is denoted as $X$ $\rightarrow$ $Y$ $\cup$ \{$e$\}. When there are multiple occurrences of the expanded event, to avoid generating too many overlapping occurrences, we restrict the expansion to only the first occurrence. Moreover, to avoid generating identical episode rules multiple times, we only expand those events that are greater than (denoted as $\succ$ in alphabetical order) the last event of $Y$. 
\end{definition}

\begin{figure}[h]
    \centering
    \includegraphics[trim=0 0 0 0,clip,scale=0.4]{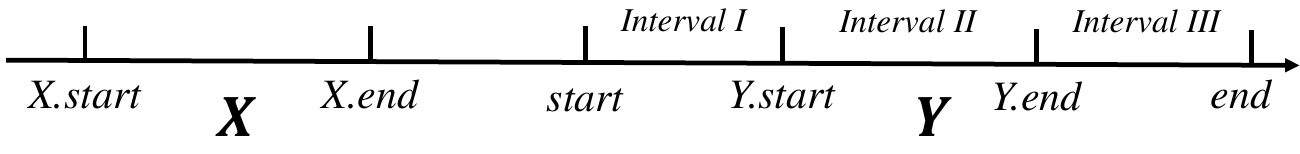}
    \caption{The \textit{search interval} of $r$ expansion \cite{chen2021mining}.}
    \label{fig:expansion}
\end{figure}

\begin{definition}[Search interval]
\label{de:search-interval}
    \rm Since the time duration constraint exists, there is also a constraint on the space of \textit{Y expansion}, called \textit{search interval} [\textit{start}, \textit{end}]. As shown in Fig. \ref{fig:expansion}, the start time stamp of the \textit{search interval} is \textit{max}($Y$.\textit{end} $-$ \textit{YSpan} + 1, $X$.\textit{end} + 1), and the end time stamp is \textit{min}($Y$.\textit{start} + \textit{YSpan} $-$ 1, $X$.\textit{end} + \textit{XYSpan} + \textit{YSpan} $-$ 2). The proof is as follows: 1) Since $Y$.\textit{start} $\geq$ $Y$.\textit{end} $-$ \textit{YSpan} + 1 and $Y$.\textit{start} $>$ $X$.\textit{end}, we can get: \textit{start} = \textit{max}($Y$.\textit{end} $-$ \textit{YSpan} + 1, $X$.\textit{end} + 1). 2) Since $Y$.\textit{start} $\leq$ $X$.\textit{end} + \textit{XYSpan} $-$ 1 and $Y$.\textit{end} $\leq$ $Y$.\textit{start} + \textit{YSpan} $-$ 1, we can get:  $Y$.\textit{end} $\leq$ $Y$.\textit{start} + \textit{YSpan} $-$ 1 $\leq$ $Y$.\textit{end} + \textit{XYSpan} + \textit{YSpan} $-$ 2 and \textit{end} = \textit{min}($Y$.\textit{start} + \textit{YSpan} $-$ 1, $X$.\textit{end} + \textit{XYSpan} + \textit{YSpan} $-$ 2).
\end{definition}

Furthermore, we analyze the update of ($Y$.\textit{start}, $Y$.\textit{end}). Suppose that $r'$: $X$ $\rightarrow$ $Y'$ is obtained from rule $r$: $X$ $\rightarrow$ $Y$ by expanding an event at $T_i$. When \textit{start} $\leq$ $i$ $<$ $Y$.\textit{start}, ($Y$.\textit{start}, $Y$.\textit{end}) will be updated to ($i$, $Y$.\textit{end}), i.e., ($Y'$.\textit{start}, $Y'$.\textit{end}). When $Y$.\textit{end} $<$ $i$ $\leq$ \textit{end}, ($Y'$.\textit{start}, $Y'$.\textit{end}) is equal to ($Y$.\textit{start}, $i$). When $Y$.\textit{start} $\leq$ $i$ $\leq$ $Y$.\textit{end}, ($Y'$.\textit{start}, $Y'$.\textit{end}) is equal to ($Y$.\textit{start}, $Y$.\textit{end}).

\begin{strategy}[Rule Event Set Pruning (RESP) strategy \cite{fournier2021mining}] \label{strategy:RESP}
    \rm If \textit{sup}($X$) is less than \textit{minsup}, then the event set $X$ cannot act as an antecedent of a promising rule, because the support of all rules that have $X$ as an antecedent is also less than \textit{minsup}. If \textit{sup}($Y$) is less than \textit{minsup} $\times$ \textit{minconf}, then the event set $Y$ cannot act as a consequent of a promising rule since the support of all rules that have $Y$ as a consequent is also less than \textit{minsup}.
\end{strategy}

\begin{definition}[Window Estimated Utilization of an event]
    \rm For an event $e$, the \textit{window estimated utilization} (\textit{WEU}) of it implies the maximum utility of the rules that contain it. The \textit{WEU} of $e$ is \textit{WEU}($e$, [$T_i$, $T_i$]) = $\sum_{k = T_i - \textit{WinSpan}}^{T_i + \textit{WinSpan}}{u(\textit{SE}_{g(k)}, k)}$, \textit{WEU}($e$) = 
    $\sum_{e \in occ(e)}{\textit{WEU}(e, [T_i, T_i])}$, where $g(k)$ denotes the time point subscript of $k$. 
\end{definition}

\begin{strategy}[WEUP strategy] \label{strategy:WEUP}
    \rm Let $e$ be an event in a complex event sequence $S$. If \textit{WEU}($e$) $<$ \textit{minutil}, which implies that the maximum utility of any rule containing $e$ is less than \textit{minutil}, then $e$ is an unpromising event and can be removed from $S$. Otherwise, it is a promising event. We call this strategy the \textit{window estimated utilization pruning (WEUP)} strategy.
\end{strategy}

\begin{definition}[Window Estimated Utilization of an episode rule]
    \rm For a rule $r$, the \textit{window estimated utilization} of it is \textit{WEU}($r$) = $\sum_{r \in occ(r)}{u(r, [T_i, T_j])}$ = $\sum_{r \in occ(r)}{\sum_{k = {T}_j-\textit{WinSpan}}^{T_i + \textit{WinSpan}}{{u}(\textit{SE}_{g(k)}, k)}}$, where the occurrence of $r$ is denoted as [$T_i$, $T_j$]. Besides, when overlapping occurrences of 1*1 rules have the same $T_i$, we take the first occurrence's \textit{WEU} value as their \textit{WEU}, because we only need to take the largest \textit{WEU} value as their \textit{WEU}. The larger the $T_j$, the smaller the \textit{WEU} value of the occurrence.
\end{definition}

\begin{figure}[h]
    \centering
    \includegraphics[trim=0 0 0 0,clip,scale=0.3]{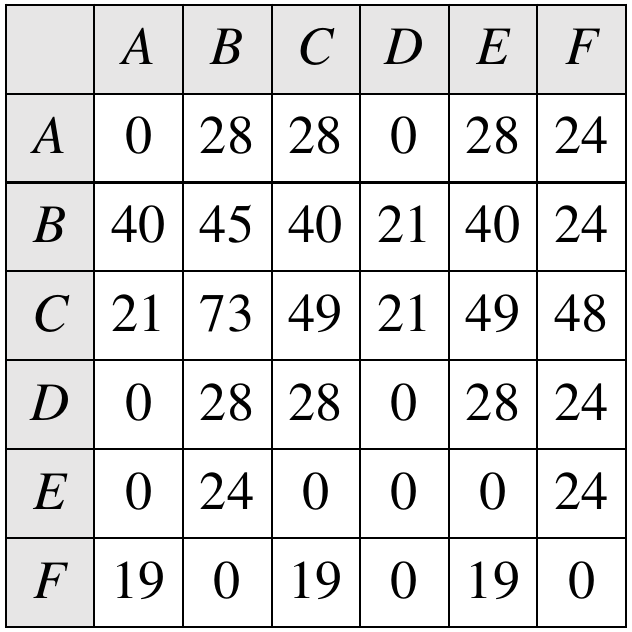}
    \caption{\textit{REUCS} when \textit{XSpan} = 1 \textit{YSpan} = 1, and \textit{XYSpan} = 3.}
    \label{fig:REUCS}
\end{figure}

If \textit{XSpan} = \textit{YSpan} = 1, \textit{XYSpan} = 3, then \textit{WinSpan} = 2, \textit{WEU}(\{$A$\}) = \textit{WEU}(\{$A$\}, [$T_2$, $T_2$]) + \textit{WEU}(\{$A$\}, [$T_6$, $T_6$]) = (0 + 4 + 7 + 10 +7) + (7 + 0 + 12 + 0 + 0) = 28 + 19 = 47, and \textit{WEU}(\{$B$\} $\rightarrow$ \{$C$\}) = \textit{WEU}(\{$B$\} $\rightarrow$ \{$C$\}, [$T_1$, $T_2$]) + \textit{WEU}(\{$B$\} $\rightarrow$ \{$C$\}, [$T_4$, $T_6$]) = (0 + 4 + 7 + 10) + (7 + 0 + 12) = 40. The \textit{Rule Estimated Utility Co-occurrence Structure (REUCS)} is in the form of a tuple ($a$, $b$, $c$), and $c$ = \textit{REUCS}($a$, $b$) = \textit{WEU}(\{$a$\} $\rightarrow$ \{$b$\}) where $a$ and $b$ are two events. As shown in Fig. \ref{fig:REUCS}, we apply a matrix to store the \textit{REUCS} values of any two events for the sequence of Fig. \ref{fig:sequence}. It can also be stored in a hash table to save memory space because there is no need to save items with a value of 0.

\begin{strategy}[REUCSP strategy] \label{strategy:REUCSP}
    \rm When a rule $r$ = $X$ $\rightarrow$ $Y$ expands an event $e$, for any event $a$ of $X$, if \textit{REUCS}($a$, $e$) $<$ \textit{minutil}, then $e$ will not perform the expansion. We call this strategy the \textit{rule estimated utility co-occurrence structure pruning (REUCSP)} strategy. The proof is as follows: Suppose that $r'$ is generated by $r$ expanding the event $e$. If the rule $\alpha$ = \{$a$\} $\rightarrow$ \{$e$\} and \textit{REUCS}($a$, $e$) $<$ \textit{minutil}, then \textit{WEU}($r'$) $\leq$ \textit{WEU}($r$) $\leq$ \textit{WEU}($\alpha$) = \textit{REUCS}($a$, $e$) $<$ \textit{minutil}, and this expansion can be terminated since $r'$ is an unpromising rule.
\end{strategy}

\begin{definition}[Rule Expansion Estimated Utilization]
    \rm Given an occurrence [$T_i$, $T_j$] of a rule $r$: $X$ $\rightarrow$ $Y$ and its \textit{search interval} [\textit{start}, \textit{end}], and the occurrence of $X$ is [$T_i$, $T_m$] where $i$ $\leq$ $m$ $<$ $j$, then the \textit{rule expansion estimated utilization (REEU)} of it is \textit{REEU}($r$, [$T_i$, $T_j$]) = $u$($X$, [$T_i$, $T_m$]) + $\sum_{k = start}^{end} {u(\textit{SE}_{g(k)}, k)}$. It represents an upper bound on the utility that the occurrence can achieve after \textit{Y expansion}.
    Besides, the \textit{rule expansion estimated utilization} of $r$ is \textit{REEU}($r$) = $\sum_{r\in nooc(r)}{\textit{REEU}(r, [T_i, T_j])}$, and $u$($r$) $\leq$ \textit{REEU}($r$). To obtain a tighter \textit{REEU}, for those overlapping occurrences, the maximum \textit{REEU} value is defined as their \textit{REEU}.
\end{definition}

For example, for a rule $r$: \{$B$\} $\rightarrow$ \{$C$\}, if \textit{XSpan} = \textit{YSpan} = \textit{XYSpan} = 3, then \textit{REEU}($r$, [$T_1$, $T_2$]) = $u$(\{$B$\}, [$T_1$, $T_1$]) + $\sum_{k=2}^{4}{u(\textit{SE}_{g(k)}, k)}$ = 26, \textit{REEU}($r$, [$T_1$, $T_3$]) = $u$(\{$B$\}, [$T_1$, $T_1$]) + $\sum_{k=2}^{5}{u(\textit{SE}_{g(k)}, k)}$ = 26, \textit{REEU}($r$, [$T_4$, $T_6$]) = $u$(\{$B$\}, [$T_4$, $T_4$]) + $\sum_{k=5}^{8}{u(\textit{SE}_{g(k)}, k)}$ = 23. Hence, \textit{REEU}($r$) = \textit{max}(\textit{REEU}($r$, [$T_1$, $T_2$]), \textit{REEU}($r$, [$T_1$, $T_3$])) + \textit{REEU}($r$, [$T_4$, $T_6$]) = 26 + 23 = 49.

\begin{theorem}[Anti-monotonicity of the REEU]
    \rm Assuming that $r'$: $X$ $\rightarrow$ $Y'$ is from the rule $r$: $X$ $\rightarrow$ $Y$ expanding the event $e$ at $T_i$. According to definition \ref{de:search-interval} and Fig. \ref{fig:expansion}, since the \textit{Y expansion} occurs only in \textit{interval I}, \textit{interval II} and \textit{interval III}, we can get $Y'$.\textit{end} $\geq$ \textit{Y.end} and $Y'$.\textit{start} $\leq$ \textit{Y.start}. Therefore, the \textit{search interval} [\textit{start}$'$, \textit{end}$'$] of $r'$ satisfies: \textit{start}$'$ $\geq$ \textit{start}, \textit{end}$'$ $\leq$ \textit{end}, and thus \textit{REEU}($r'$) $\leq$ \textit{REEU}($r$).
\end{theorem}
 
\begin{strategy}[REEUP strategy]
    \rm Let [\textit{start}, \textit{end}] be the \textit{search interval} of a candidate rule $r$. We have $u$($r$) $\leq$ \textit{REEU}($r$). If \textit{REEU}($r$) $<$ \textit{minutil}, then $r$ is not a HUER, and we can terminate the subsequent expansion of it. This is the rule expansion estimated utilization pruning (REEUP) strategy.
\end{strategy}

\subsection{Data Structure}

\begin{figure}[h]
    \centering
    \includegraphics[trim=0 0 0 0,clip,scale=0.32]{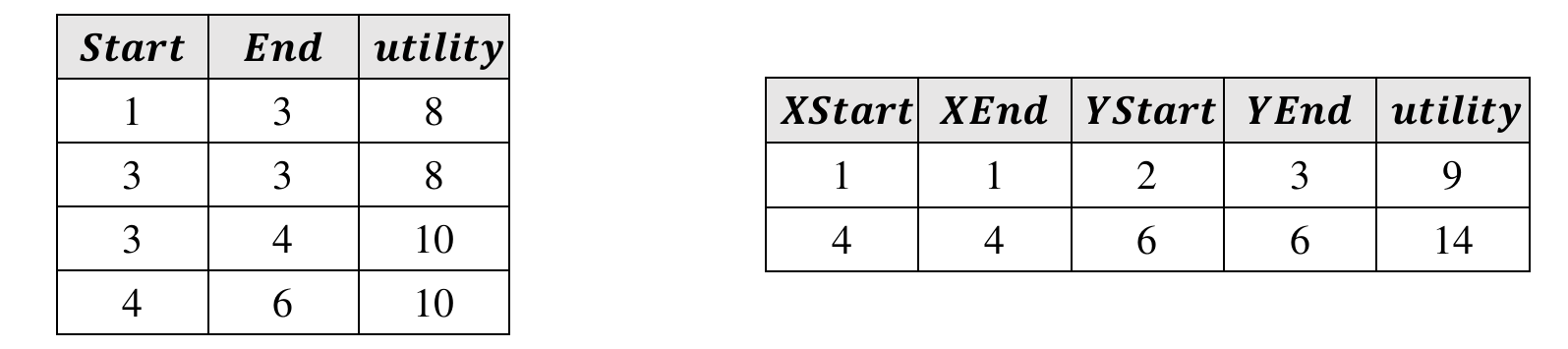}
    \caption{The \textit{NoList} of $F$: \{$B$, $E$\} (left) and $r$: \{$B$\} $\rightarrow$ \{$C$, $E$\} (right) when \textit{XSpan} = 3, \textit{YSpan} = 2 and \textit{XYSpan} = 3.}
    \label{fig:NoList}
\end{figure}

UPER uses a list named \textit{NoList} to store all occurrences of an event set $F$ or a rule $r$. It records information about the interval and utility of each occurrence. For an occurrence \textit{No} of $F$, it is in the form of a tuple (\textit{start}, \textit{end}, \textit{utility}), while a \textit{No} of $r$ is in the form of a tuple (\textit{XStart}, \textit{XEnd}, \textit{YStart}, \textit{YEnd}, \textit{utility}). \textit{NoList} has several advantages. The utility is the sum of the \textit{utility} of all non-overlapping occurrences in \textit{NoList}, and the number of non-overlapping occurrences in \textit{NoList} is the \textit{support} of $F$ or $r$. In addition, the information of the interval can facilitate the calculation of the upper bound \textit{REEU} of rules. According to Definition \ref{de:expansion}, when \textit{No} is expanded, the associated interval and utility may change along with it; \textit{YStart}, \textit{YEnd}, and \textit{utility} in \textit{NoList} will update. As shown in Fig. \ref{fig:NoList}, there are two instances of the \textit{NoList}.

\subsection{UPER Algorithm}

\begin{figure}[h]
    \centering
    \includegraphics[trim=0 0 0 0,clip,scale=0.38]{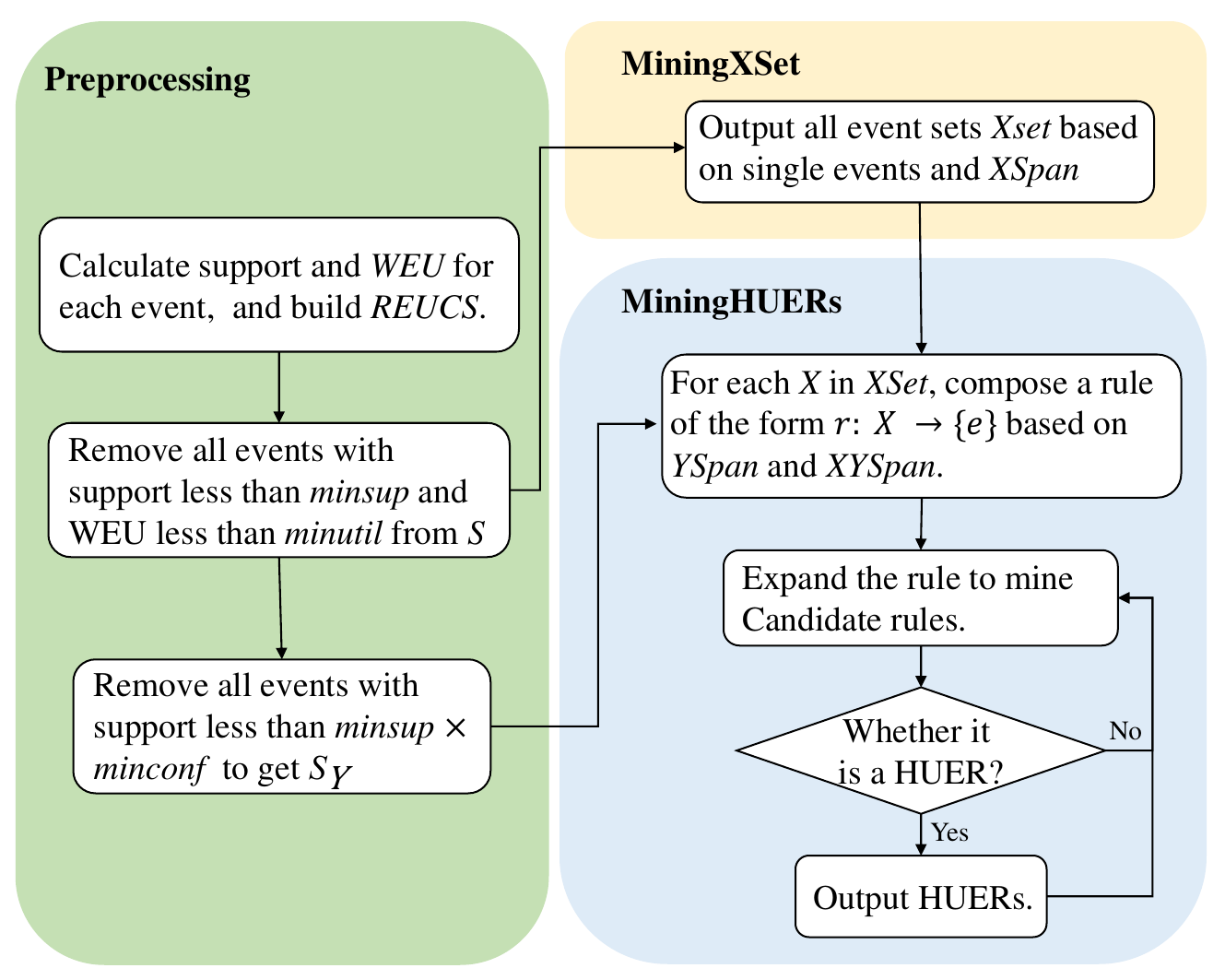}
    \caption{The framework of UPER.}
    \label{fig:framework}
\end{figure}

Here, we describe the running process of UPER in detail, and then the framework of UPER is shown in Fig. \ref{fig:framework}. 

\vspace{-5pt} 
\begin{algorithm}[!h]
    \small
    \caption{UPER}
    \label{alg:UPER}
    \LinesNumbered
    \KwIn{a complex event sequence: $S$, the minimum utility: \textit{minutil}, the minimum support: \textit{minsup}, the minimum confidence: \textit{minconf}, the time duration constraint: \textit{XSpan}, \textit{YSpan}, and \textit{XYSpan}.}
    \KwOut{the set of \textit{HUERs}.}

    scan $S$ to 1). calculate the \textit{support} of each $e$ $\in$ $E$; 2). calculate the \textit{WEU} of each $e$ $\in$ $E$; 3). build \textit{REUCS}; \\
\For{\rm each event $e$ $\in$ $E$}{
    \If{\textit{sup}($e$) $<$ \textit{minsup} \rm{or} \textit{WEU}($e$) $<$ \textit{minutil}}{
        remove $e$ from $E$ and $S$;\\  // \textbf{the RESP strategy, the WEUP strategy}
    }
}
\textit{XSet} $\leftarrow$ call \textbf{MiningXSet}($S$, \textit{XSpan}, \textit{minsup}); \\
call \textbf{MiningHUERs}($S$, \textit{XSet}, \textit{XSpan}, \textit{XYSpan}, \textit{YSpan}, \textit{minsup}, \textit{minconf}, \textit{minutil});\\
\end{algorithm}

\vspace{-5pt} 
\begin{algorithm}[!h]
    \small
    \caption{MiningXSet}
    \label{alg:miningXset}
    \LinesNumbered
    \KwIn{a complex event sequence: $S$, \textit{XSpan}, \textit{minsup}.}
    \KwOut{a list of event sets with utility that may be antecedents of \textit{HUERs}.}

scan $S$ to build a map \textit{EventSetMap} $<$key = \{$e$\}, value = \textit{NoList}$>$ for each event $e$; \\
\textit{XSet} $\leftarrow$ remove the pairs from \textit{EventSetMap} whose $|$\textit{NoList}$|$ $<$ \textit{minsup}; \ \ // \textbf{the RESP strategy}\\
\For{\rm each event set $F$ in \textit{XSet}}{
    \textit{NoList} $\leftarrow$ the occurrence list of $F$; \\
    clear \textit{EventSetMap};\\
    \For{\rm each occurrence \textit{No} in \textit{NoList}}{
        \textit{start} $\leftarrow$ \textit{No}.\textit{end} $-$ \textit{XSpan} + 1;\\
        \textit{end} $\leftarrow$ \textit{No}.\textit{start} + \textit{XSpan} $-$ 1;\\
        \For{\rm each event $e$ in [\textit{start}, \textit{end}]}{
            \If{$e$ $\succ$ $F$.\textit{lastEvent}}{
                $F$$'$ $\leftarrow$ $F$ $\cup$ \{$e$\};\\
                \textit{No}$'$ $\leftarrow$ record the occurrence of $F$$'$;\\
                add $<$$F$$'$, \textit{No}$'$$>$ into \textit{EventSetMap};\\
            }
        }
        add each pair of \textit{EventSetMap} whose $|$\textit{NoList$'$}$|$ $\geq$ \textit{minsup} into \textit{XSet};
    }
}
\Return{\textit{XSet}}
\end{algorithm}

\vspace{-5pt} 
\begin{algorithm}[h]
    \small
    \caption{MiningHUERs}
    \label{alg:miningHUERs}
    \LinesNumbered
    \KwIn{a complex sequence: $S$, a list of event sets that may be antecedents of \textit{HUERs}: \textit{XSet}, thresholds: \textit{XSpan}, \textit{XYSpan}, \textit{YSpan}, \textit{minsup}, \textit{minconf}, \textit{minutil}.}
    \KwOut{the set of \textit{HUERs}.}
    
    $S_Y$ $\leftarrow$ remove all events with support less than \textit{minsup} $\times$ \textit{minconf} from $S$;
    \ \ \ // \textbf{the RESP strategy}\\
\For{\rm each event set $X$ in \textit{XSet}}{
    \textit{rNoList}, \textit{rNoList}$'$, \textit{CandidateQueue} $\leftarrow$ $\emptyset$;\\
    \For{\rm each occurrence \textit{XNo} in \textit{X.NoList}}{
        \For{\rm each event $e$ in [\textit{XNo}.\textit{end} + 1, \textit{XNo}.\textit{end} + \textit{YSpan} + \textit{XYSpan}] of $S_Y$}{
            \For{each event $a$ in $X$}{
            \If{\textit{REUCS}($a$, $e$) $<$ \textit{minutil}}{
                continue; \ \ // \textbf{the REUCSP strategy} \\
            }
            }
         \textit{rNo} $\leftarrow$ record the occurrence of the rule $X$ $\rightarrow$ \{$e$\};\\
        }
        add \textit{rNo} to \textit{rNoList};\\
    }
    \If{\textit{REEU(r)} $<$ \textit{minutil}}{
        continue;\ \ \ \  // \textbf{the REEUP strategy}
    }
    add $<$$r$, \textit{rNoList}$>$ to \textit{CandidateQueue}; \\
    \If{$u$($r$) $\geq$ \textit{minutil} \rm{and} \textit{sup}($r$) $\geq$ \textit{minconf} $\times$ \textit{sup}($X$)}{
        ouput $r$;
    }
    \While{CandidateQueue $\neq$ $\emptyset$}{
        $r$ $\leftarrow$ pop a rule $X$ $\rightarrow$ $Y$ from \textit{CandidateQueue};\\
        \For{\rm each occurrence \textit{rNo} in \textit{r.NoList}}{
            calculate the \textit{search interval} [\textit{start}, \textit{end}];\\
            \For{\rm each event $e$ in [\textit{start}, \textit{end}] of $S_Y$ and $e$ $\succ$ \textit{Y.LastEvent}}{
            \For{each event $a$ in $X$}{
            \If{\textit{REUCS}($a$, $e$) $<$ \textit{minutil}}{
                continue; \ \ // \textbf{the REUCSP strategy} \\
            }
            }
            \textit{rNo}$'$ $\leftarrow$ record the occurrence of the rule $X$ $\rightarrow$ $Y$ $\cup$ \{$e$\};\\
            }
            add \textit{rNo}$'$ to \textit{rNoList}$'$;\\
        }
        \If{\textit{REEU}($r'$) $<$ \textit{minutil}}{
            continue; \ \ \ // \textbf{the REEUP strategy}
        }
        \If{$u$($r'$) $\geq$ \textit{minutil} \rm{and} \textit{sup}($r'$) $\geq$ \textit{minconf} $\times$ \textit{sup}($X$)}{
            ouput $r'$;\\
        }
        add $<$$r'$, \textit{rNoList}$'$$>$ to \textit{CandidateQueue};\\
    }
}
\end{algorithm}

As is shown in the pseudocode of Algorithm \ref{alg:UPER}, UPER is divided into three procedures: preprocessing, then mining the set of events \textit{XSet} that may be the antecedents of HUERs, and finally mining all HUERs based on the $X$ in \textit{XSet}. The input parameters of UPER are a complex event sequence $S$ and user-specified thresholds, including \textit{XSpan}, \textit{YSpan}, \textit{XYSpan}, \textit{minutil}, \textit{minsup}, and \textit{minconf}. At the preprocessing procedure, UPER first scans $S$ to calculate the \textit{support} and \textit{WEU} of distinct events and builds a hashmap to store the REUCS values of any two events (line 1). Based on the WEUP strategy and RESP strategy, those events that can not be the antecedents of a HUER will be removed (lines 2-7).

Subsequently, UPER calls \textit{MiningXSet} to discover a list of event sets that may serve as antecedents of HUERs (line 8), as presented in Algorithm \ref{alg:miningXset}. \textit{MiningXSet} first scans $S$ to build the \textit{NoList} for each event set consisting of a single event $e$, storing the information in a map named \textit{XSet} in the form of $<$\{$e$\}, \textit{NoList}$>$ (line 1). Based on the RESP strategy, the procedure removes those events that cannot act as rule antecedents and saves the remaining events in the \textit{XSet} (line 2). Then the expansion is performed on the event sets in \textit{XSet} one by one to get all the event sets that can act as rule antecedents (lines 3-18). For the start position \textit{start} of the expansion interval, it takes a minimum value of (\textit{No.end} $-$ \textit{XSpan} + 1) since \textit{No.end} $-$ start $\leq$ \textit{XSpan} $-$ 1. Similarly, the maximum value of the end position of the expansion interval is (\textit{No.start} + \textit{XSpan} $-$ 1) (lines 7-8). In addition, to avoid getting a duplicate event set, the procedure only expands events that are greater than the last event of the current event set (line 10).

After the execution of \textit{MiningXSet}, UPER calls \textit{MiningHUERs} to discover all valid HUERs (line 9), which is presented in Algorithm \ref{alg:miningHUERs}. Based on the RESP strategy, it first removes all events with support less than \textit{minsup} $\times$ \textit{minconf} from $S$ to obtain $S_Y$ (line 1). Then, each event set in the \textit{XSet} is traversed through a loop to find its consequents and compose the rule (lines 2-44). In the loop, the procedure first identifies single events from $S_Y$ to serve as consequents and forms a rule $r$: $X$ $\rightarrow$ \{$e$\} (lines 3-14). For each item $a$ in $X$, if \textit{REUCS(a, e)} $<$ \textit{minutil}, then $r$ and its expansion results are not high utility (the REUCSP strategy). Additionally, the \textit{REEU} $r$ is computed by the \textit{search interval} of its occurrences (lines 15-17). If \textit{REEU(r)} is less than \textit{minutil}, the subsequent expansion (the REEUP strategy) can be terminated. Otherwise, $r$ is stored in the candidate queue \textit{CandidateQueue} and it is determined whether it is a HUER (lines 18-21). Then the procedure expands all rules in \textit{CandidateQueue} to obtain the new rule $r'$ (lines 22-43). It is worth noting that when \textit{REEU($r'$)} is less than \textit{minutil}, there is no need to expand $r'$ (REEUP strategy). Otherwise, we also determine if it is an HUER and store $r'$ to \textit{CandidateQueue} (lines 39-42).

\begin{figure*}[h]
    \centering
    \includegraphics[trim=0 0 0 0,clip,scale=0.5]{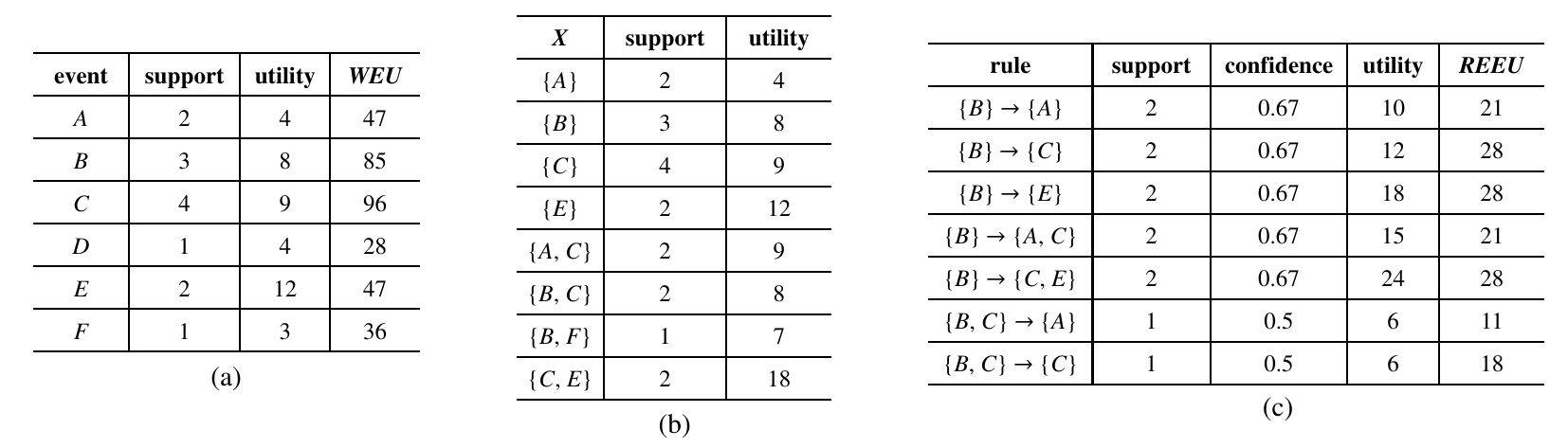}
    \caption{Some details of the example.}
    \label{fig:example_algo}
\end{figure*}

To describe the algorithm more clearly, here we take the sequence in Fig. \ref{fig:sequence} as an example. When \textit{minsup} = 2, \textit{minconf} = 0.6, \textit{minutil} = 20, \textit{XSpan} = 1, \textit{YSpan} = 1, and \textit{XYSpan} = 3, we show some details of the three procedures of the algorithm in Fig. \ref{fig:example_algo}. Fig. \ref{fig:example_algo} (a) shows the information for all distinct events after preprocessing, including support, utility, and \textit{WEU}. Since the support of both $D$ and $F$ is less than \textit{minsup}, they can be removed from $S$ according to the RESP strategy. However, all of them have values of \textit{WEU} greater than \textit{minutil}, which means that we cannot rely on the WEUP strategy for pruning. Fig. \ref{fig:example_algo} (b) shows all event sets that may be the antecedent of a rule after the MiningXSet procedure. Their specific information also includes the start and end positions, stored in their respective \textit{NoList}, which we have omitted here. In Fig. \ref{fig:example_algo} (c), there are several candidate rules related to event $B$. Among them, only the utility of \{$B$\} $\rightarrow$ \{$C$, $E$\} is no less than \textit{minutil}, and it has a support and confidence greater than the thresholds, i.e., only it is HUER. Furthermore, notice that both \{$B$, $C$\} $\rightarrow$ \{$A$\} and \{$B$, $C$\} $\rightarrow$ \{$C$\} have values of \textit{REEU} less than \textit{minutil}, we do not continue to expand on them according to the REEUP strategy. According to Fig. \ref{fig:REUCS} and the \textit{REUCSP} strategy, \{$F$\} $\rightarrow$ \{$A$\} cannot be a promising rule because \textit{REUCS}($F$, $A$) = 19 $<$ \textit{minutil}.

\section{Experiments} \label{sec: experiments}

In this section, we evaluate the effect of UPER from several perspectives. Note that UPER is the first algorithm to consider utility in partially-ordered episode rule mining, and there are no suitable baseline algorithms for comparison because the different mining tasks are not suitable for comparison. Hence, we designed several variants of UPER to evaluate the effectiveness of different pruning strategies.

\subsection{Datasets and Experimental Setup}

\textbf{Datasets.} In this experiment, we conducted evaluations of time and memory consumption on six real-life datasets. All datasets are in the form of a complex event sequence, where each event has its own utility. They are available on the SPMF website\footnote{\url{http://www.philippe-fournier-viger.com/spmf}}, and the details of them are shown in Table \ref{tab:dataset}. Since these datasets are widely used in experiments with utility-driven episode mining algorithms \cite{wu2013mining, fournier2019hue, gan2023discovering}, they are also applicable to the UPER algorithm. Besides, it is worth noting that while these datasets may be regarded as a transaction dataset, they can be regarded as a complex event sequence when each item is considered as an event and each transaction as a simultaneous event set \cite{fournier2021mining}.

\begin{table}[!h]
    \footnotesize
    \centering
    \caption{Characteristics of the datasets}
    \label{tab:dataset}
    \begin{tabular}{|c|c|c|c|}
\hline
\textbf{Dataset}          & \textbf{Time point} & \textbf{\# of events} & \textbf{Average length} \\ \hline
\textbf{Retail}           & 88,162              & 16,470                & 10.3                    \\ \hline
\textbf{Kosarak}          & 990,002             & 41,270                & 8.1                     \\ \hline
\textbf{Chainstore}       & 1,112,949           & 46,086                & 7.23                    \\ \hline
\textbf{Foodmart}         & 4,141               & 1,559                 & 4.42                    \\ \hline
\textbf{Yoo-choose-buy}         & 234,300               & 16,004                   & 2.165                      \\ \hline
\textbf{Ecommerce retail} & 14,975              & 3,468                 & 11.71                   \\ \hline
\end{tabular}
\end{table}

\textbf{Experimental setup.} Our experiments are performed on a PC equipped with a 64-bit Windows 10 operating system, a 3.6 GHz AMD Ryzen 5 3600 CPU, and 32 GB of RAM. UPER is implemented in Java, and the runtime and memory consumption are measured using the Java API. For reproducibility, all the source code and datasets are available at GitHub\footnote{\url{https://github.com/DSI-Lab1/UPER}}.

\textbf{Algorithm variants.} We develop UPER into four variants (UPER$_1$, UPER$_2$, UPER$_3$, and UPER$_4$), the difference among them is the adoption of pruning strategies REUCSP and REEUP. Table \ref{tab:variants} demonstrates the adoption of the pruning strategy by different variants of UPER in detail.

\begin{table}[!ht]
    \small
    \centering
    \caption{Different variants of UPER}
    \label{tab:variants}
    \begin{tabular}{c|c|c}
    \hline
    \textbf{Variants} & \textbf{REUCSP strategy} & \textbf{REEUP strategy} \\ \hline
    UPER$_1$          & $\times$                 & $\times$                \\ \hline
    UPER$_2$          & $\checkmark$                 & $\times$            \\ \hline
    UPER$_3$          & $\times$             & $\checkmark$                \\ \hline
    UPER$_4$          & $\checkmark$             & $\checkmark$            \\ \hline
    \end{tabular}
\end{table}

\subsection{Impact of the Utility Threshold on Runtime}

\begin{figure*}[h]
    \centering
   \includegraphics[trim=0 0 0 0,clip,width=18cm, height=7.7cm]{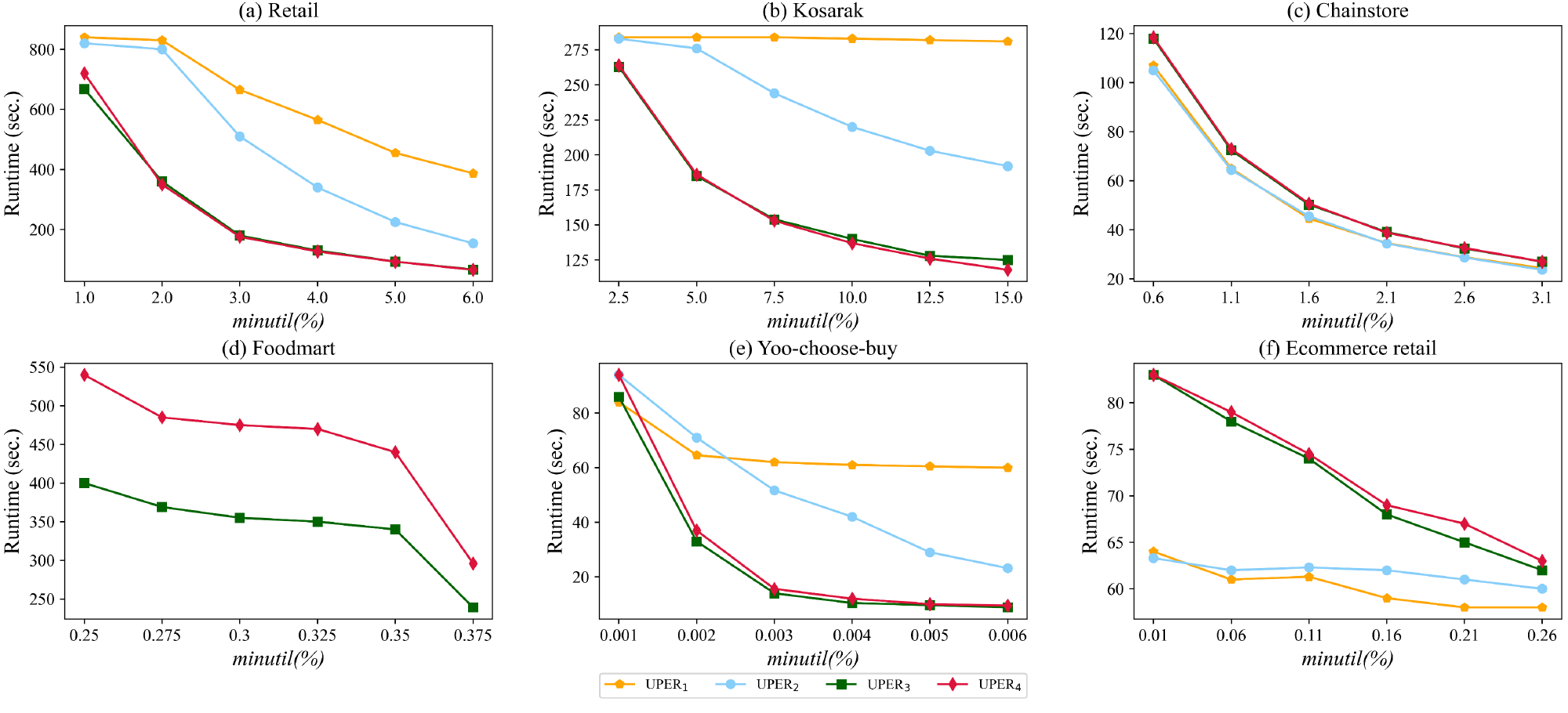}
    \caption{Runtime of different algorithm variants.}
    \label{fig:utility}
\end{figure*}

We first evaluate the effect of different utility thresholds in UPER variants. Since different datasets have their own characteristics, other thresholds should be changed along with them. Table \ref{tab:parameter} shows the specific parameter settings. Fig. \ref{fig:utility} presents the running time of variants on different datasets. Additionally, we evaluate the number of candidate rules generated by the different variants, as shown in Table \ref{tab:candidate}.

\begin{table}[h]
    \centering
    \footnotesize
    \caption{Parameter settings}
    \label{tab:parameter}
    \resizebox{\linewidth}{!}{ 
\begin{tabular}{c|c|c|c|c|c}
\hline
\textbf{Dataset}          & \textit{\textbf{minsup}} & \textit{\textbf{minconf}} & \textit{\textbf{XSpan}} & \textit{\textbf{YSpan}} & \textit{\textbf{XYSpan}} \\ \hline
\textbf{Retail}           & 100                      & 0.4                                            & 2                                            & 4                                            & 4                        \\ \hline
\textbf{Kosarak}          & 10000                    & 0.4                                            & 2                                            & 3                                            & 3                        \\ \hline
\textbf{Chainstore}       & 250                      & 0.3                                            & 2                                            & 4                                            & 4                        \\ \hline
\textbf{Foodmart}         & 2                        & 0.3                                            & 2                                            & 4                                            & 4                        \\ \hline
\textbf{Yoo-choose-buy}   & 3                        & 0.2                                            & 2                                            & 4                                            & 4                        \\ \hline
\textbf{Ecommerce retail} & 20                       & 0.3                                            & 3                                            & 4                                            & 4                        \\ \hline
\end{tabular}}
\end{table}

\begin{table*}[h]
    \centering
    \caption{Number of candidates under different utility thresholds}
    \includegraphics[trim=0 0 0 0,clip,scale=0.33]{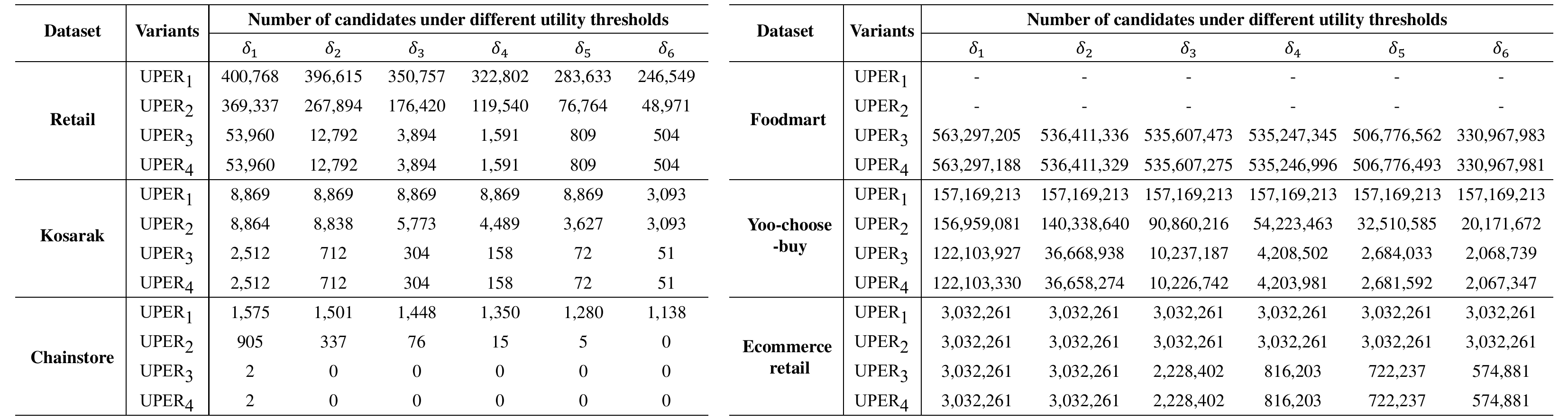}
    \label{tab:candidate}
\end{table*}

\begin{figure*}[h]
    \centering
    \includegraphics[trim=0 0 0 0,clip,width=18cm, height=7.7cm]{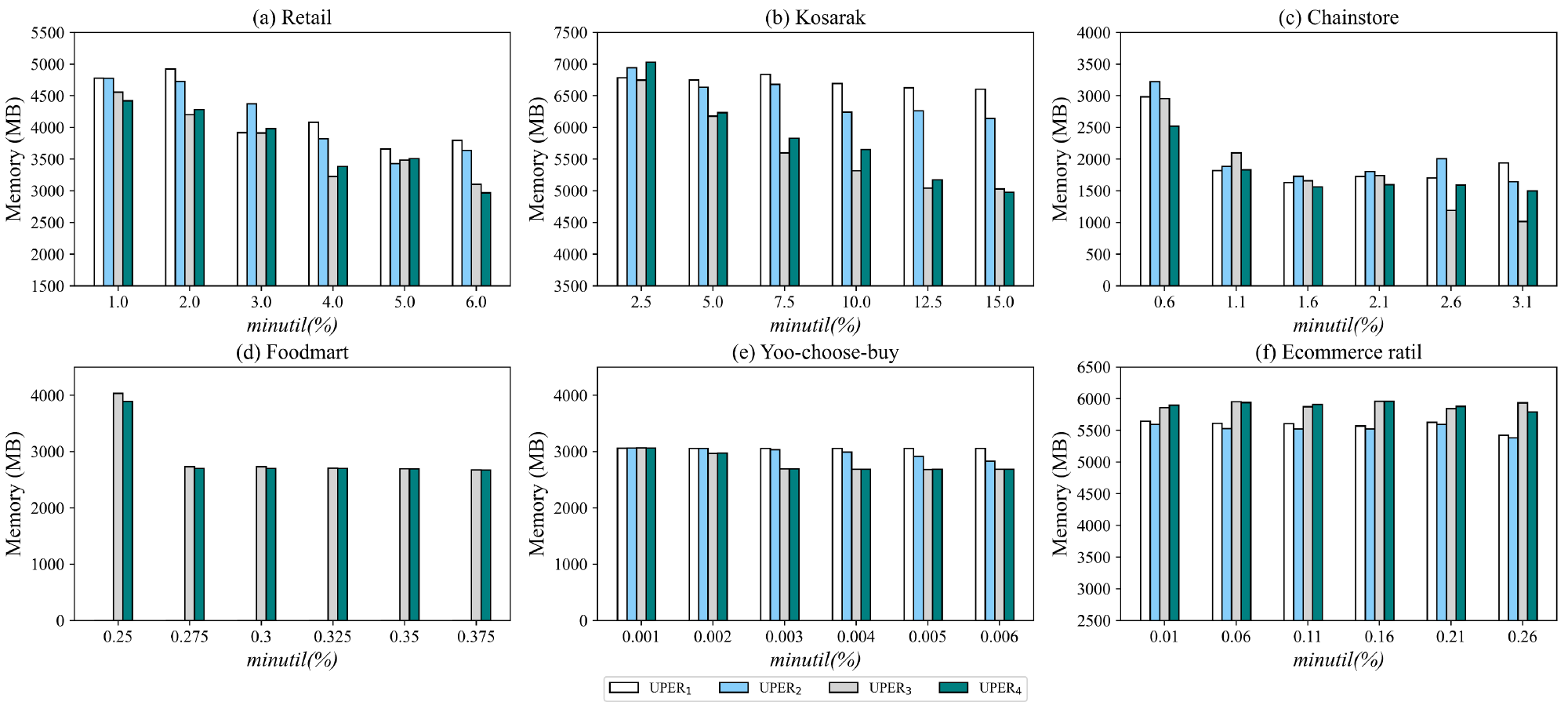}
    \caption{Memory consumption of different algorithm variants.}
    \label{fig:memory}
\end{figure*}

On the datasets Retail, Kosarak, and Yoo-choose-buy, UPER$_3$ and UPER$_4$ are much faster than UPER$_{1}$ and UPER$_{2}$, which means that the REEUP strategy effectively reduces the search space. It can also be seen in Table \ref{tab:candidate} that UPER$_3$ and UPER$_4$ generate 10 times fewer candidate rules than UPER$_1$ and UPER$_2$. Additionally, as the \textit{minutil} increases, UPER$_2$ will save more time than UPER$_1$, while UPER$_4$ is not much different from UPER$_3$. That is, when UPER does not use the tighter utility upper bound, the more effective the REUCSP strategy will be as the \textit{minutil} increases. Furthermore, when the \textit{minutil} is low, UPER$_1$ may be slower than UPER$_2$. For example, on Yoo-choose-buy, it takes time to query the REUCS while the pruning space is not large. Besides, when UPER uses the tighter utility upper bound, the REUCSP strategy does not have an advantage since the search space has been reduced. It can be seen in Table \ref{tab:candidate} that the number of candidate rules generated by UPER$_3$ and UPER$_4$ is close in most cases. On the Foodmart dataset, it's worth noting that UPER$_1$ and UPER$_2$ are not shown in the figure because they both run for more than 10 minutes. We can find that UPER$_3$ is faster than UPER$_4$. This is because the values in REUCS are generally larger, resulting in a limited REUCSP strategy. For the datasets Chainstore and Ecommerce retail, it is interesting that UPER$_3$ and UPER$_4$ are slower than UPER$_1$ and UPER$_2$, and we analyze the possible reasons below in the candidate number and dataset characteristics. For the Chainstore dataset, although the REEUP strategy reduces the number of candidate rules, the time spent on computing \textit{REEU} increases because the time points of Chainstore are large. For the Ecommerce retail dataset with a large average length, the number of events and the average utility of all time points are also larger, which will result in a larger value of REEU for most rules, and the number of candidates pruned by the REEUP strategy is 0 when the \textit{minutil} is low. As the \textit{minutil} increases, the REEUP strategy can prune a certain number of candidate rules, and the time spent by UPER$_3$ and UPER$_4$ begins to be similar to that of UPER$_1$ and UPER$_2$.

\subsection{Impact of the Utility Threshold on Memory Cost}

Moreover, as is shown in Fig. \ref{fig:memory}, we also evaluate the memory consumption of UPER variants. On the retail dataset, UPER$_1$ and UPER$_2$ consume more memory than UPER$_3$ and UPER$_4$ in most cases. The reason is that the REEUP strategy saves memory by pruning redundant candidate rules. Similarly, we can notice that UPER$_1$ and UPER$_2$ generally consume more memory than UPER$_3$ and UPER$_4$, except on the Ecommerce retail dataset. In the Ecommerce retail dataset, the REEUP strategy is not effective in reducing the number of candidate rules, as can be seen from the runtime, and instead consumes memory to compute the upper bound REEU of the rules. Furthermore, we also focus on the memory consumption of the UPER variants with or without the REUCSP strategy. On the Retail and Kosarak datasets, since the REUCSP strategy is more effective in UPER variants that do not adopt the REEU strategy, i.e., UPER$_1$ and UPER$_2$, UPER$_2$ consumes less memory than UPER$_1$, while the opposite is true for UPER$_3$ and UPER$_4$. For the same reason, the difference in memory consumed by UPER$_1$ and UPER$_2$ is, in general, larger than that of UPER$_3$ and UPER$_4$. On the Chainstore dataset, the values are large in most REUCS, hence the REUCSP strategy doesn't work to its advantage when the \textit{minutil} is low. This is reflected in the fact that UPER$_2$ does not save as much memory as UPER$_1$, and in some cases consumes more; the same phenomenon occurs in UPER$_3$ and UPER$_4$. On the Foodmart and Yoo-choose-buy datasets, UPER$_3$ and UPER$_4$ cost a little difference in memory. The reason is that the average length of these two datasets is small, so there are not many events for rule expansion, and querying REUCS consumes very little memory.

\subsection{Impact of XSpan}

\begin{figure*}[h]
    \centering
    \includegraphics[trim=0 0 0 0,clip,width=18cm, height=4.3cm]{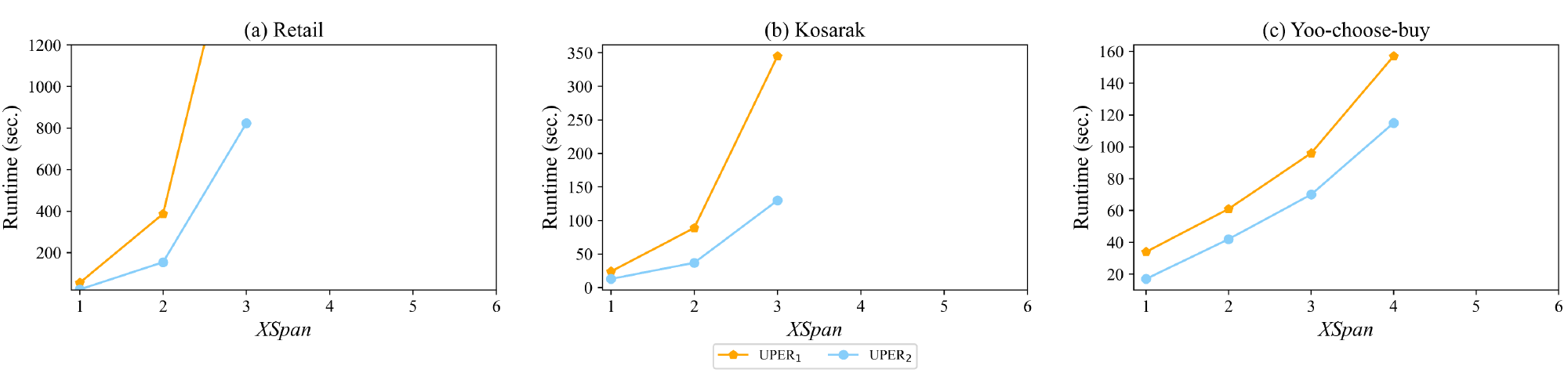}
    \caption{The impact of \textit{XSpan} w.r.t the REUCSP strategy.}
    \label{fig:xspan}
\end{figure*}

Since the REUCSP strategy requires scanning every event in the rule antecedent, we believe that it is worthwhile to observe the impact of the \textit{XSpan} on it. For ease of observation, we select three datasets, i.e., the Retail, Kosarak, and Yoo-choose-buy. The algorithm variants involved in this experiment are UPER$_1$ and UPER$_2$. The parameter settings are in Table \ref{tab:XSpan} and the experiment results are shown in Fig. \ref{fig:xspan}. We also count the number of candidate rules generated by UPER$_1$ and UPER$_2$ in Table \ref{fig:XSpan_candidate}. It is obvious that as \textit{XSpan} increases, UPER$_2$ is getting much faster than UPER$_1$. The main reason is that the larger the \textit{XSpan} is, the number of antecedent events of a rule will increase, hence the number of candidate rules available for pruning by the REUCSP strategy will increase, and the speed of UPER$_2$ will get ahead of UPER$_1$. In short, assuming that a REUCS has enough values that are less than \textit{minutil}, then as \textit{XSpan} increases, the more effective the REUCSP strategy is.

\begin{table}[h]
    \centering
    \footnotesize
    \caption{Parameter settings for the experiment w.r.t. \textit{XSpan}}
    \label{tab:XSpan}
    \resizebox{\linewidth}{!}{ 
\begin{tabular}{c|c|c|c|c|c}
\hline
\textbf{Dataset}        & \textit{\textbf{minsup}} & \multicolumn{1}{l|}{\textit{\textbf{minconf}}} & \multicolumn{1}{l|}{\textit{\textbf{minutil (\%)}}} & \multicolumn{1}{l|}{\textit{\textbf{YSpan}}} & \textit{\textbf{XYSpan}} \\ \hline
\textbf{Retail}         & 100                      & 0.4                                            & 6.0                                                & 4                                            & 4                        \\ \hline
\textbf{Kosarak}        & 15000                    & 0.4                                            & 65.0                                               & 3                                            & 3                        \\ \hline
\textbf{Yoo-choose-buy} & 3                        & 0.2                                            & 0.004                                              & 4                                            & 4                        \\ \hline
\end{tabular}}
\end{table}

\begin{table*}[h]
    \centering
    \caption{Number of candidates under different \textit{XSpan}}
    \includegraphics[trim=0 0 0 0,clip,scale=0.42]{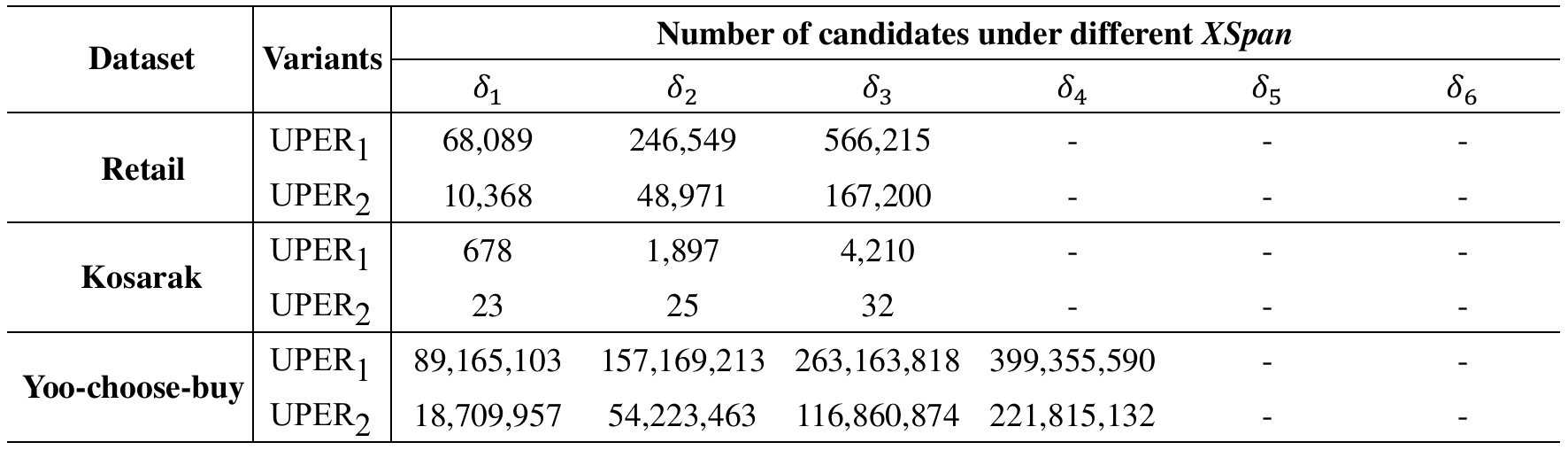}
    \label{fig:XSpan_candidate}
\end{table*}

\begin{figure*}[!ht]
    \centering
    \includegraphics[trim=0 0 0 0,clip,width=18cm, height=7.7cm]{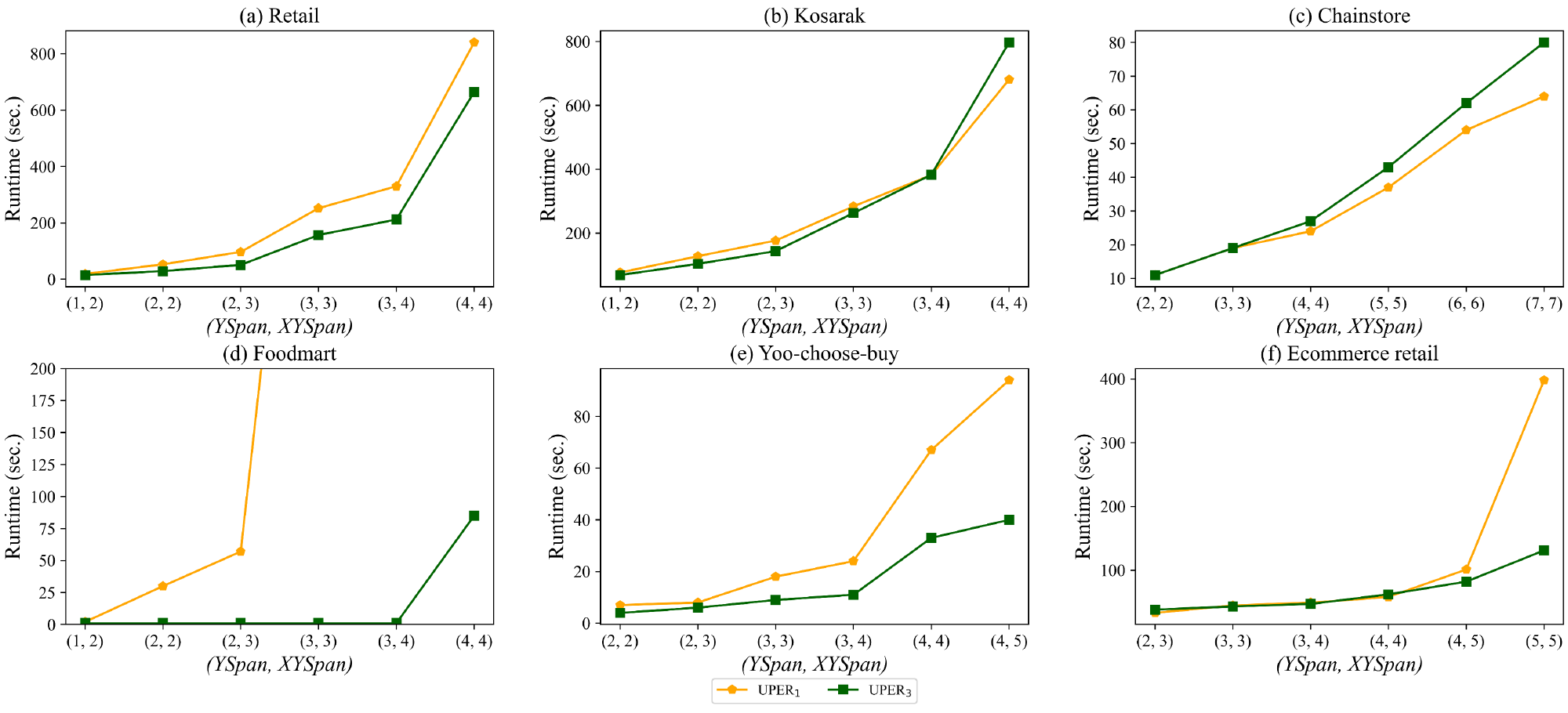}
    \caption{The impact of \textit{YSpan} and \textit{XYSpan} w.r.t. the REEUP strategy.}
    \label{fig:YSpan}
\end{figure*}

\begin{table*}[!ht]
    \centering
    \caption{Number of candidates under different \textit{YSpan} and \textit{XYSpan}}
    \includegraphics[trim=0 0 0 0,clip,scale=0.42]{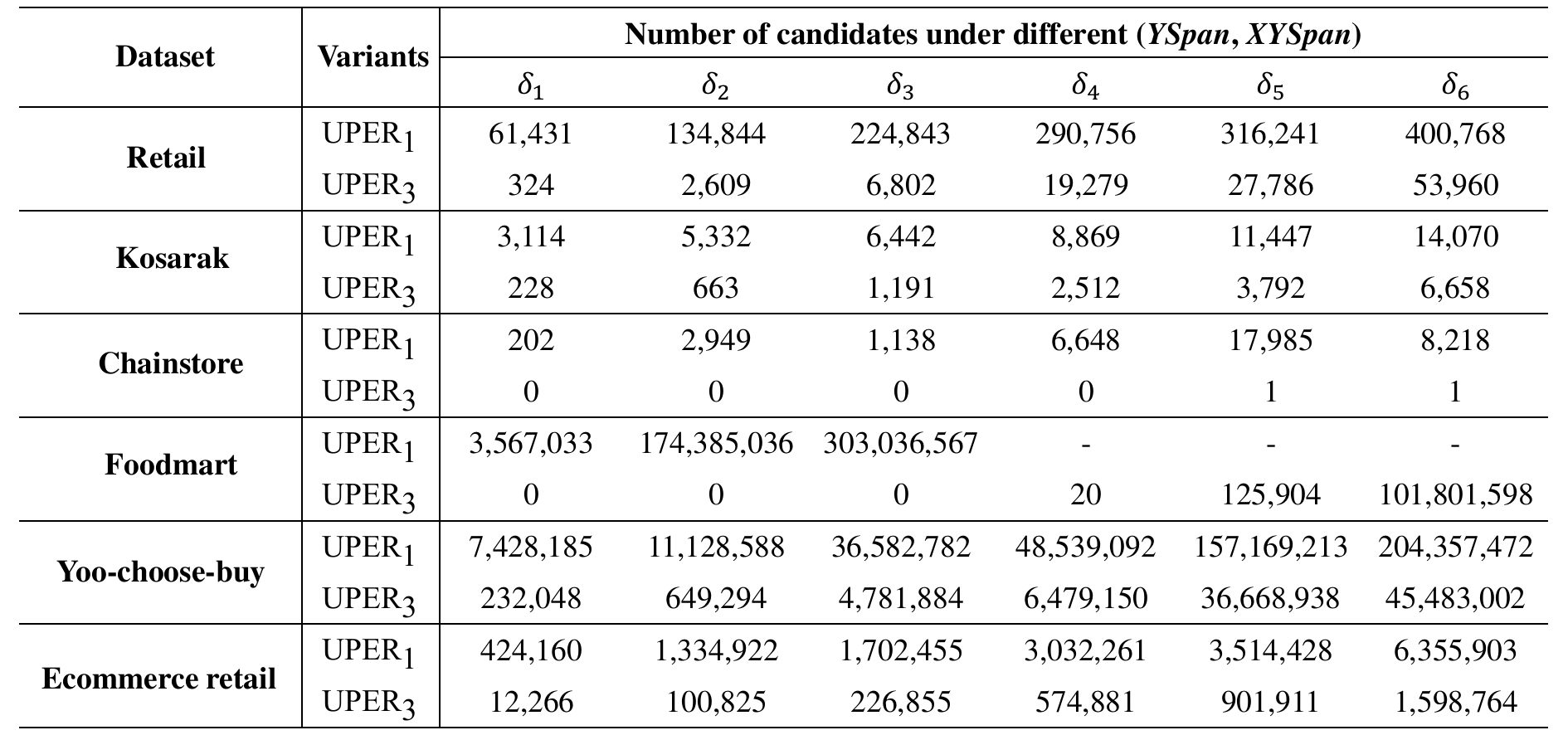}
    \label{fig:YSpan_candidate}
\end{table*}

\subsection{Impact of YSpan and XYSpan}

The expansion events of the rule come from the \textit{search interval}, and the size of \textit{search interval} depends on the \textit{XYSpan} and \textit{YSpan}. Here, we explore the impact of the \textit{XYSpan} and \textit{YSpan} on the REEUP strategy. The parameter settings are in Table \ref{tab:YSpan}, and the experiment results are shown in Fig. \ref{fig:YSpan}. The algorithm variants involved in this experiment are UPER$_1$ and UPER$_3$. Besides, we also count the number of candidate rules generated by UPER$_1$ and UPER$_3$ in Table \ref{fig:YSpan_candidate}. The main regularity we can observe is that as \textit{YSpan} and \textit{XYSpan} increase, UPER$_3$ is getting much faster than UPER$_1$, e.g., on the Retail, Foodmart, Yoo-choose-buys, and Ecommerce retail datasets. This comes from a simple principle that when \textit{YSpan} and \textit{XYSpan} are small, UPER$_1$, which does not use the REEUP strategy, will not spend much time on rule expansion. When \textit{YSpan} and \textit{XYSpan} increase, UPER$_3$ could save more time on rule expansion than UPER$_1$ by removing numerous candidate rules. However, the remaining two datasets do not seem to follow this regularity. On the Kosarak dataset, UPER$_3$ is first tied with UPER$_1$, then ahead of UPER$_1$, and finally overtaken by UPER$_1$ and gradually falling behind. On the Chainstore dataset, UPER$_3$ shows a little difference from UPER$_1$, and is increasingly slower than UPER$_1$. This is because, for both datasets, the utility thresholds are set lower. As \textit{YSpan} and \textit{XYSpan} increase, the length of the \textit{serach interval} will become longer, which makes the value of \textit{REEU} too high, and the REEUP strategy will lose its advantage, and so UPER$_3$ will be gradually slower than UPER$_1$.

In conclusion, when \textit{YSpan} and \textit{XYSpan} are small, although the REEUP strategy can reduce some candidate rules, it makes little difference whether the REEUP strategy is adopted or not, because the length of \textit{search interval} is small and there is no need to scan many time points for rule expansion. As \textit{YSpan} and \textit{XYSpan} increase, the REEUP strategy is effective in removing candidate rules and saving the time spent on rule expansion. However, if both \textit{YSpan} and \textit{XYSpan} can increase continuously, the length of the \textit{search interval} will be too long, which results in overestimation of the \textit{REEU} and REEUP strategies losing their advantage.

\begin{table}[h]
    \centering
    \caption{Parameter settings for the experiment w.r.t. \textit{YSpan} and \textit{XYSpan}}
    \label{tab:YSpan}
    \begin{tabular}{c|c|c|c|c}
    \hline
    \textbf{Dataset}          & \textit{\textbf{minsup}} & \textit{\textbf{minconf}} & \textit{\textbf{minutil(\%)}} & \textit{\textbf{XSpan}} \\ \hline
    \textbf{Retail}           & 100                      & 0.4                       & 1.0                           & 2                       \\ \hline
    \textbf{Kosarak}          & 10000                    & 0.4                       & 2.5                           & 2                       \\ \hline
    \textbf{Chainstore}       & 250                      & 0.3                       & 3.1                           & 2                       \\ \hline
    \textbf{Foodmart}         & 2                        & 0.3                       & 0.4                           & 2                       \\ \hline
    \textbf{Yoo-choose-buy}   & 3                        & 0.2                       & 0.002                         & 2                       \\ \hline
    \textbf{Ecommerce retail} & 20                       & 0.3                       & 0.26                          & 3                       \\ \hline
    \end{tabular}
\end{table}

\section{Conclusion}  \label{sec: conclusion}

In this paper, we consider utility-driven mining of partially ordered episode rules and propose the UEPR algorithm. We first define the utility of POERs and a data structure named \textit{NOList} to store information about all non-overlapping occurrences for an event set or a rule, including position and utility. Besides, we propose the \textit{search interval} to describe the region of rule expansion and analyze in detail the changes in the \textit{search interval} before and after the rule expansion. Additionally, to improve the efficiency of the algorithm, we propose two utility upper bounds, namely window estimated utilization (\textit{WEU}) and rule expansion estimated utilization (\textit{REEU}). Based on them, three pruning strategies, WEUP, REUCSP, and REEUP, are proposed. Finally, we perform several experiments to explore the impact of the utility threshold and time duration constraints (i.e., \textit{XSpan}, \textit{YSpan}, and \textit{XYSpan}) on pruning strategies, and we analyze the possible reasons and summarize the regularity. In the future, we will consider how to increase the running speed or reduce the memory consumption of the algorithm. On the one hand, similar to the POERM algorithm, a rule's expansion is performed only after all the possible antecedents of the rule have been stored, which is more demanding in terms of memory. Therefore, improving the framework of the algorithm may address this limitation. On the other hand, it is also interesting to explore a tighter upper bound when the value of the time duration constraint is large. In addition, since POERM has been shown to achieve better prediction results \cite{chen2021sequence}, we will consider applying UPER to real predictions to verify that it can predict more meaningful information. For example, we can discover more information rather than just frequent smartphone events, as indicated in the study \cite{goyal2023smartphone}. It is also useful to apply UPER to website intrusion detection, stock investment prediction, and similar tasks.

\bibliographystyle{IEEEtran}
\bibliography{paper}

@inproceedings{fournier2021mining,
  title={Mining partially-ordered episode rules in an event sequence},
  author={Fournier-Viger, Philippe and Chen, Yangming and Nouioua, Farid and Lin, Jerry Chun-Wei},
  booktitle={13th Asian Conference on Intelligent Information and Database Systems},
  pages={3--15},
  year={2021},
  organization={Springer}
}

@inproceedings{chen2021mining,
  title={Mining partially-ordered episode rules with the head support},
  author={Chen, Yangming and Fournier-Viger, Philippe and Nouioua, Farid and Wu, Youxi},
  booktitle={23rd International Conference on Big Data Analytics and Knowledge Discovery},
  pages={266--271},
  year={2021},
  organization={Springer}
}

@article{lin2015discovering,
  title={Discovering utility-based episode rules in complex event sequences},
  author={Lin, Yu-Feng and Wu, Cheng-Wei and Huang, Chien-Feng and Tseng, Vincent S},
  journal={Expert Systems with Applications},
  volume={42},
  number={12},
  pages={5303--5314},
  year={2015},
  publisher={Elsevier}
}

@inproceedings{fournier2022pattern,
  title={Pattern mining: Current challenges and opportunities},
  author={Fournier-Viger, Philippe and Gan, Wensheng and Wu, Youxi and Nouioua, Mourad and Song, Wei and Truong, Tin and Duong, Hai},
  booktitle={International Conference on Database Systems for Advanced Applications},
  pages={34--49},
  year={2022},
  organization={Springer}
}

@article{gan2019survey,
  title={A survey of parallel sequential pattern mining},
  author={Gan, Wensheng and Lin, Jerry Chun Wei and Fournier-Viger, Philippe and Chao, Han-Chieh and Yu, Philip S},
  journal={ACM Transactions on Knowledge Discovery from Data},
  volume={13},
  number={3},
  pages={1--34},
  year={2019},
  publisher={ACM}
}

@article{gan2023discovering,
  title={Discovering High Utility Episodes in Sequences},
  author={Gan, Wensheng and Lin, Jerry Chun Wei and Chao, Han Chieh and Yu, Philip S},
  journal={IEEE Transactions on Artificial Intelligence},
  volume={4},
  number={03},
  pages={473--486},
  year={2023},
  publisher={IEEE}
}

@article{wu2017nosep,
  title={{NOSEP}: Nonoverlapping sequence pattern mining with gap constraints},
  author={Wu, Youxi and Tong, Yao and Zhu, Xingquan and Wu, Xindong},
  journal={IEEE Transactions on Cybernetics},
  volume={48},
  number={10},
  pages={2809--2822},
  year={2017},
  publisher={IEEE}
}

@article{wu2020netncsp,
  title={{NetNCSP}: Nonoverlapping closed sequential pattern mining},
  author={Wu, Youxi and Zhu, Changrui and Li, Yan and Guo, Lei and Wu, Xindong},
  journal={Knowledge-Based Systems},
  volume={196},
  pages={105812},
  year={2020},
  publisher={Elsevier}
}

@article{ouarem2023discovering,
  title={Discovering frequent parallel episodes in complex event sequences by counting distinct occurrences},
  author={Ouarem, Oualid and Nouioua, Farid and Fournier-Viger, Philippe},
  journal={Applied Intelligence},
  pages={1--7},
  year={2023},
  publisher={Springer}
}

@article{ao2019large,
  title={Large-scale frequent episode mining from complex event sequences with hierarchies},
  author={Ao, Xiang and Shi, Haoran and Wang, Jin and Zuo, Luo and Li, Hongwei and He, Qing},
  journal={ACM Transactions on Intelligent Systems and Technology},
  volume={10},
  number={4},
  pages={1--26},
  year={2019},
  publisher={ACM}
}

@article{baek2021rhups,
  title={{RHUPS}: Mining recent high utility patterns with sliding window--based arrival time control over data streams},
  author={Baek, Yoonji and Yun, Unil and Kim, Heonho and Nam, Hyoju and Kim, Hyunsoo and Lin, Jerry Chun-Wei and Vo, Bay and Pedrycz, Witold},
  journal={ACM Transactions on Intelligent Systems and Technology},
  volume={12},
  number={2},
  pages={1--27},
  year={2021},
  publisher={ACM}
}

@article{alam2023ugmine,
  title={{UGMINE}: utility-based graph mining},
  author={Alam, Md Tanvir and Roy, Amit and Ahmed, Chowdhury Farhan and Islam, Md Ashraful and Leung, Carson K},
  journal={Applied Intelligence},
  volume={53},
  number={1},
  pages={49--68},
  year={2023},
  publisher={Springer}
}

@article{fournier2019efficient,
  title={Efficient algorithms to identify periodic patterns in multiple sequences},
  author={Fournier-Viger, Philippe and Li, Zhitian and Lin, Jerry Chun-Wei and Kiran, Rage Uday and Fujita, Hamido},
  journal={Information Sciences},
  volume={489},
  pages={205--226},
  year={2019},
  publisher={Elsevier}
}

@article{gan2021survey,
  title={A survey of utility-oriented pattern mining},
  author={Gan, Wensheng and Lin, Jerry Chun Wei and Fournier-Viger, Philippe and Chao, Han Chieh and Tseng, Vincent S and Yu, Philip S},
  journal={IEEE Transactions on Knowledge and Data Engineering},
  volume={33},
  number={4},
  pages={1306--1327},
  year={2021},
  publisher={IEEE}
}

@article{gan2021fast,
  title={Fast utility mining on sequence data},
  author={Gan, Wensheng and Lin, Jerry Chun Wei and Zhang, Jiexiong and Fournier-Viger, Philippe and Chao, Han Chieh and Yu, Philip S},
  journal={IEEE Transactions on Cybernetics},
  volume={51},
  number={2},
  pages={487--500},
  year={2021},
  publisher={IEEE}
}

@inproceedings{fournier2019hue,
  title={{HUE-Span}: Fast high utility episode mining},
  author={Fournier-Viger, Philippe and Yang, Peng and Lin, Jerry Chun-Wei and Yun, Unil},
  booktitle={Advanced Data Mining and Applications},
  pages={169--184},
  year={2019},
  organization={Springer}
}

@inproceedings{wu2013mining,
  title={Mining high utility episodes in complex event sequences},
  author={Wu, Cheng-Wei and Lin, Yu-Feng and Yu, Philip S and Tseng, Vincent S},
  booktitle={ACM SIGKDD International Conference on Knowledge Discovery and Data Mining},
  pages={536--544},
  year={2013}
}

@inproceedings{chen2021sequence,
  title={Sequence prediction using partially-ordered episode rules},
  author={Chen, Yangming and Fournier-Viger, Philippe and Nouioua, Farid and Wu, Youxi},
  booktitle={International Conference on Data Mining Workshops},
  pages={574--580},
  year={2021},
  organization={IEEE}
}

@inproceedings{meger2004constraint,
  title={Constraint-based mining of episode rules and optimal window sizes},
  author={M{\'e}ger, Nicolas and Rigotti, Christophe},
  booktitle={European Conference on Principles of Data Mining and Knowledge Discovery},
  pages={313--324},
  year={2004},
  organization={Springer}
}

@inproceedings{fournier2022maxfem,
  title={{MaxFEM}: Mining maximal frequent episodes in complex event sequences},
  author={Fournier-Viger, Philippe and Nawaz, M Saqib and He, Yulin and Wu, Youxi and Nouioua, Farid and Yun, Unil},
  booktitle={International Conference on Multi-disciplinary Trends in Artificial Intelligence},
  pages={86--98},
  year={2022},
  organization={Springer}
}

@article{laxman2005discovering,
  title={Discovering frequent episodes and learning hidden markov models: A formal connection},
  author={Laxman, Srivatsan and Sastry, PS and Unnikrishnan, KP},
  journal={IEEE Transactions on Knowledge and Data Engineering},
  volume={17},
  number={11},
  pages={1505--1517},
  year={2005},
  publisher={IEEE}
}

@article{mannila1997discovery,
  title={Discovery of frequent episodes in event sequences},
  author={Mannila, Heikki and Toivonen, Hannu and Inkeri Verkamo, A},
  journal={Data Mining and Knowledge Discovery},
  volume={1},
  pages={259--289},
  year={1997},
  publisher={Springer}
}

@article{huang2008efficient,
  title={Efficient mining of frequent episodes from complex sequences},
  author={Huang, Kuo-Yu and Chang, Chia-Hui},
  journal={Information Systems},
  volume={33},
  number={1},
  pages={96--114},
  year={2008},
  publisher={Elsevier}
}

@inproceedings{fournier2020tke,
  title={{TKE}: Mining top-$k$ frequent episodes},
  author={Fournier-Viger, Philippe and Yang, Yanjun and Yang, Peng and Lin, Jerry Chun-Wei and Yun, Unil},
  booktitle={33rd International Conference on Industrial, Engineering and Other Applications of Applied Intelligent Systems},
  pages={832--845},
  year={2020},
  organization={Springer}
}

@article{fournier2017survey,
  title={A survey of sequential pattern mining},
  author={Fournier-Viger, Philippe and Lin, Jerry Chun-Wei and Kiran, Rage Uday and Koh, Yun Sing and Thomas, Rincy},
  journal={Data Science and Pattern Recognition},
  volume={1},
  number={1},
  pages={54--77},
  year={2017}
}

@article{gan2021huopm,
  title={{HUOPM}: High-utility occupancy pattern mining},
  author={Gan, Wensheng and Lin, Jerry Chun-Wei and Fournier-Viger, Philippe and Chao, Han-Chieh and Yu, Philip S},
  journal={IEEE Transactions on Cybernetics},
  volume={50},
  number={3},
  pages={1195--1208},
  year={2020},
  publisher={IEEE}
}

@article{zhang2023mining,
  title={Mining high-utility sequences with positive and negative values},
  author={Zhang, Xiaojie and Lai, Fuyin and Chen, Guoting and Gan, Wensheng},
  journal={Information Sciences},
  volume={637},
  pages={118945},
  year={2023},
  publisher={Elsevier}
}

@inproceedings{guo2014high,
  title={High utility episode mining made practical and fast},
  author={Guo, Guangming and Zhang, Lei and Liu, Qi and Chen, Enhong and Zhu, Feida and Guan, Chu},
  booktitle={Advanced Data Mining and Applications},
  pages={71--84},
  year={2014},
  organization={Springer}
}

@article{huang2023us,
  title={{US-Rule}: Discovering utility-driven sequential rules},
  author={Huang, Gengsen and Gan, Wensheng and Weng, Jian and Yu, Philip S},
  journal={ACM Transactions on Knowledge Discovery from Data},
  volume={17},
  number={1},
  pages={1--22},
  year={2023},
  publisher={ACM}
}

@article{ao2017mining,
  title={Mining precise-positioning episode rules from event sequences},
  author={Ao, Xiang and Luo, Ping and Wang, Jin and Zhuang, Fuzhen and He, Qing},
  journal={IEEE Transactions on Knowledge and Data Engineering},
  volume={30},
  number={3},
  pages={530--543},
  year={2017},
  publisher={IEEE}
}

@inproceedings{ouarem2021mining,
  title={Mining episode rules from event sequences under non-overlapping frequency},
  author={Ouarem, Oualid and Nouioua, Farid and Fournier-Viger, Philippe},
  booktitle={International Conference on Industrial, Engineering and Other Applications of Applied Intelligent Systems},
  pages={73--85},
  year={2021},
  organization={Springer}
}

@inproceedings{fournier2014erminer,
  title={{ERMiner}: sequential rule mining using equivalence classes},
  author={Fournier-Viger, Philippe and Gueniche, Ted and Zida, Souleymane and Tseng, Vincent S},
  booktitle={Advances in Intelligent Data Analysis},
  pages={108--119},
  year={2014},
  organization={Springer}
}

@article{jarvela2023predicting,
  title={Predicting regulatory activities for socially shared regulation to optimize collaborative learning},
  author={J{\"a}rvel{\"a}, Sanna and Nguyen, Andy and Vuorenmaa, Eija and Malmberg, Jonna and J{\"a}rvenoja, Hanna},
  journal={Computers in Human Behavior},
  volume={144},
  pages={107737},
  year={2023},
  publisher={Elsevier}
}

@inproceedings{goyal2023smartphone,
  title={Smartphone Context Event Sequence Prediction with {POERMH} and {TKE-Rules} Algorithms},
  author={Goyal, Pooja and Khan, Md Khorrom and Steil, Christian and Martel, Sarah M and Bryce, Renee},
  booktitle={Annual Computing and Communication Workshop and Conference},
  pages={0827--0834},
  year={2023},
  organization={IEEE}
}

@article{lin2017novel,
  title={A novel methodology for stock investment using high utility episode mining and genetic algorithm},
  author={Lin, Yu-Feng and Huang, Chien-Feng and Tseng, Vincent S},
  journal={Applied Soft Computing},
  volume={59},
  pages={303--315},
  year={2017},
  publisher={Elsevier}
}

@article{nguyen2023efficient,
  title={An efficient method for mining high occupancy itemsets based on equivalence class and early pruning},
  author={Nguyen, Loan TT and Mai, Thang and Pham, Giao-Huy and Yun, Unil and Vo, Bay},
  journal={Knowledge-Based Systems},
  volume={267},
  pages={110441},
  year={2023},
  publisher={Elsevier}
}

@article{kim2022ehmin,
  title={{EHMIN}: Efficient approach of list based high-utility pattern mining with negative unit profits},
  author={Kim, Heonho and Ryu, Taewoong and Lee, Chanhee and Kim, Hyeonmo and Yoon, Eunchul and Vo, Bay and Lin, Jerry Chun-Wei and Yun, Unil},
  journal={Expert Systems with Applications},
  volume={209},
  pages={118214},
  year={2022},
  publisher={Elsevier}
}

@inproceedings{rathore2016top,
  title={Top-$k$ high utility episode mining from a complex event sequence},
  author={Rathore, Sonam and Dawar, Siddharth and Goyal, Vikram and Patel, Dhaval},
  booktitle={International Conference on Management of Data},
  pages={56--62},
  year={2016}
}

@article{amiri2018online,
  title={An online learning model based on episode mining for workload prediction in cloud},
  author={Amiri, Maryam and Mohammad-Khanli, Leyli and Mirandola, Raffaela},
  journal={Future Generation Computer Systems},
  volume={87},
  pages={83--101},
  year={2018},
  publisher={Elsevier}
}

@inproceedings{agrawal1995mining,
  title={Mining sequential patterns},
  author={Agrawal, Rakesh and Srikant, Ramakrishnan},
  booktitle={International Conference on Data Engineering},
  pages={3--14},
  year={1995},
  organization={IEEE}
}

@inproceedings{fournier2014fast,
  title={Fast vertical mining of sequential patterns using co-occurrence information},
  author={Fournier-Viger, Philippe and Gomariz, Antonio and Campos, Manuel and Thomas, Rincy},
  booktitle={Advances in Knowledge Discovery and Data Mining},
  pages={40--52},
  year={2014},
  organization={Springer}
}

@inproceedings{yin2012uspan,
  title={{USpan}: an efficient algorithm for mining high utility sequential patterns},
  author={Yin, Junfu and Zheng, Zhigang and Cao, Longbing},
  booktitle={ACM SIGKDD International Conference on Knowledge Discovery and Data Mining},
  pages={660--668},
  year={2012}
}

@article{lan2014applying,
  title={Applying the maximum utility measure in high utility sequential pattern mining},
  author={Lan, Guo-Cheng and Hong, Tzung-Pei and Tseng, Vincent S and Wang, Shyue-Liang},
  journal={Expert Systems with Applications},
  volume={41},
  number={11},
  pages={5071--5081},
  year={2014},
  publisher={Elsevier}
}

@inproceedings{fournier2011rulegrowth,
  title={{RuleGrowth}: mining sequential rules common to several sequences by pattern-growth},
  author={Fournier-Viger, Philippe and Nkambou, Roger and Tseng, Vincent Shin-Mu},
  booktitle={ACM Symposium on Applied Computing},
  pages={956--961},
  year={2011}
}

@inproceedings{zida2015efficient,
  title={Efficient mining of high-utility sequential rules},
  author={Zida, Souleymane and Fournier-Viger, Philippe and Wu, Cheng-Wei and Lin, Jerry Chun-Wei and Tseng, Vincent S},
  booktitle={Machine Learning and Data Mining in Pattern Recognition},
  pages={157--171},
  year={2015},
  organization={Springer}
}

\end{document}